\documentclass[12pt]{article}

\usepackage{epsf}
\usepackage{latexsym,amssymb,euscript}
\usepackage[dvips]{graphicx}
\usepackage{amsmath}

\begin{document}
\begin{titlepage}

\begin{center}
{\Large \textbf{ The next-to-leading order jet vertex for Mueller-Navelet and
forward jets revisited }}
\end{center}

\centerline{F.~Caporale$^{1\dagger}$, D.Yu.~Ivanov$^{2\P}$,
B.~Murdaca$^{1\dagger}$, A.~Papa$^{1\dagger}$, A.~Perri$^{1\dagger}$}

\vskip .6cm

\centerline{${}^1$ {\sl Dipartimento di Fisica, Universit\`a della Calabria,}}
\centerline{\sl and Istituto Nazionale di Fisica Nucleare, Gruppo collegato di
Cosenza,}
\centerline{\sl I-87036 Arcavacata di Rende, Cosenza, Italy}

\vskip .2cm

\centerline{${}^2$ {\sl Sobolev Institute of Mathematics and
Novosibirsk State University,}}
\centerline{\sl 630090 Novosibirsk, Russia}

\vskip 2cm

\begin{abstract}
We recalculate,  within the  BFKL approach and at the
next-to-leading order, the jet vertex relevant for the production of
Muel\-ler-Na\-velet jets in proton collisions and of forward jets in DIS.
We consider both processes with incoming quark and gluon. The starting point
is the definition of quark and gluon impact factors in the BFKL approach.
Following this procedure we show explicitly
that all infrared divergences cancel when renormalized parton densities are
considered. We compare our results for the vertex with the former calculation
of Refs.~\cite{bar1,bar2}.
\end{abstract}

%\vskip .5cm

$
\begin{array}{ll}
^{\dagger}\mbox{{\it e-mail address:}} &
\mbox{francesco.caporale, beatrice.murdaca, alessandro.papa,}\\
& \mbox{amedeo.perri\ @fis.unical.it}\\
^{\P}\mbox{{\it e-mail address:}} &
\mbox{d-ivanov@math.nsc.ru}\\
\end{array}
$

\end{titlepage}

\newpage

\section{Introduction}
\label{intro}

The Mueller-Navelet jet production process~\cite{Mueller:1986ey} was suggested
as an ideal tool to study the Regge limit of perturbative Quantum
ChromoDynamics (QCD) in proton-proton (or proton-antiproton) collisions.
It is an inclusive process
\begin{equation}
p(p_1)+p(p_2)\to J_1(k_{J,1})+J_2(k_{J,2})+X \, ,
\label{process}
\end{equation}
where two hard jets $J_1$ and $J_2$ are produced (the transverse momenta of
jets are much larger than the QCD scale, $\vec k_{J,1}^2\sim \vec k_{J,2}^2
\gg \Lambda_{\rm QCD}^2$) in the kinematical region where the jets are
separated by a large interval of rapidity, $\Delta y\gg 1$.
This regime requires large center of mass energy $s$ of the proton
collisions, $s=2p_1\cdot p_2\gg \vec k_{J \, 1,2}^2$, since
$\Delta y\sim\ln {s/\vec k^2_{J\, 1,2}}$. It can be studied experimentally
at modern high energy hadron colliders, LHC and Tevatron.

The BFKL approach~\cite{BFKL} is the most suitable framework for the
theoretical description of the high-energy limit of hard or semi-hard
processes. At high $s$, large logarithms of the energy
compensate the small coupling and must be resummed at all orders of the
perturbative series. The BFKL approach provides a systematic way to
perform the resummation in the leading logarithmic approximation (LLA), which
means resummation of all terms $(\alpha_s\ln(s))^n$, and in the
next-to-leading logarithmic approximation (NLA), which means resummation of
all terms $\alpha_s(\alpha_s\ln(s))^n$.

In the BFKL approach, both in the LLA and in the NLA, the high-energy
scattering amplitudes are expressed by a suitable factorization of a
process-independent part, the Green's function for the interaction of two
Reggeized gluons, and process-dependent terms, the so-called impact factors
of the colliding particles~(see, for instance,~\cite{FF98}).
The Green's function is determined through the BFKL equation, whose kernel is
known at the next-to-leading order (NLO)~\cite{FL98,CC98}.
The impact factors of the colliding particle are a necessary ingredient
for the complete description of a process in the BFKL approach and, therefore,
to get a contact with phenomenology. The only impact factors calculated
so far with NLO accuracy are those for colliding quark and
gluons~\cite{fading,fadinq,Cia,Ciafaloni:2000sq}, for forward jet 
production~\cite{bar1,bar2},
for the $\gamma^* \to \gamma^*$ transition~\cite{gammaIF}, and for the
$\gamma^*$ to light vector meson transition at leading twist~\cite{IKP04}.

The D0 collaboration at Tevatron~\cite{Abbott:1999ai} observed power-like rise
of the Mueller-Navelet jet cross section with energy, though the D0 results
revealed an even stronger rise than predicted by LLA BFKL calculations.
Besides the cross section measurement, it was suggested to study a less
inclusive observable, such as the decorrelation of jets in the relative
azimuthal angle. The D0 experiment~\cite{Abachi:1996et} found less
decorrelation than predicted by LLA
calculations~\cite{DelDuca:1993mn,Stirling:1994zs}.

Important improvements were made toward the description of the process within
NLA accuracy. Effects related with QCD running coupling were studied
in~\cite{Orr:1997im,Kwiecinski:2001nh}. 
In~\cite{Vera:2006un,Vera:2007kn,Marquet:2007xx}
the full NLO Green's function was implemented, but the jets impact factors
were taken into account at the leading order only.

Recently the results of a complete NLA analysis of the process~(\ref{process})
were reported~\cite{Colferai:2010wu}, which incorporates NLO corrections to
both the BFKL Green's function and the jets impact factors, calculated earlier
in~\cite{bar1,bar2}.
The authors of~\cite{Colferai:2010wu} found that for kinematics typical of the
LHC experiment the effect of NLO corrections to the jet impact factors is very
important, of the same order as the one obtained from the NLO correction to
Green's function. This observation is similar to one obtained earlier in the
NLA analysis of the diffractive double $\rho$-electroproduction~\cite{mesons}.
Another important conclusion of~\cite{Colferai:2010wu} is that the results for
Mueller-Navelet jet observables obtained within complete BFKL NLA analysis
appeared to be very close to the one calculated in the conventional collinear
factorization at the NLO, with the only exception of the ratio between
the azimuthal angular moments $\langle \cos(2\phi) \rangle /\langle \cos \phi
\rangle$.

In our opinion it would be important to have an independent calculation of
Mueller-Navelet jet observables in NLA. The aim of the present paper
is the calculation of NLO correction to the jet impact factor in order to have
an independent check of the results of~\cite{bar1,bar2}.
In many technical steps we follow closely the method used in~\cite{bar1,bar2},
but we will take advantage of starting from the general definition
for the impact factors at NLO, see~\cite{FF98}, which allows us to come to the
results more shortly.

The paper is organized as follows. In the next Section we will present the
factorization structure of the cross section, recall the definition
of BFKL impact factor and discuss the treatment of the divergences arising
in the calculation; in Section~3 we give the derivation of the quark
contribution to the impact factor; Section~4 is devoted to the calculation of
the gluon part; finally, in Section~5 we summarize our results and make a
comparison to the ones of~\cite{bar1,bar2}.

\section{General framework}
\label{frame}

The state of the jets can be described completely by their rapidities,
$y_{1,2}$, and transverse momenta, $\vec k_{J,1}$ and  $\vec k_{J,2}$.
We denote the azimuthal angles of the produced jets as $\phi_{1}$ and
$\phi_{2}$. It is convenient to define the Sudakov decomposition for the jets
momenta. For a jet in the fragmentation region of the proton with momentum
$p_1$, one has
\begin{equation}
k_{J,1}= x_{J,1} p_1+ \frac{\vec k^2_{J,1}}{x_{J,1} s}p_2+k_{J,1 \, \perp} \ ,
\quad\quad\quad k_{J,1\, \perp}^2=-\vec k^2_{J,1} \ ,
\end{equation}
where we assume $p_1^2=p_2^2=0$ neglecting the proton mass and the
longitudinal fraction $x_{J,1}={\cal O}(1)$ is related to the jet rapidity in
the center of mass system by
$$y_1=\frac{1}{2}\ln\frac{x_{J,1}^2 s}{\vec k_{J,1}^2}\;,\quad\quad
dy_1=\frac{dx_{J,1}}{x_{J,1}}\; .$$

In QCD collinear factorization the cross section of the process reads
\begin{equation}
\frac{d\sigma}{dx_{J,1}dx_{J,2}d^2\vec k_{J,1}d^2\vec k_{J,2}}
=\sum_{i,j=q,\bar q,g}\int\limits^1_0
\int\limits^1_0 dx_1dx_2 f_i(x_1,\mu) f_j(x_2,\mu)
\frac{d\hat \sigma_{i,j}(x_1 x_2 s,\mu)}{dx_{J,1}dx_{J,2}d^2\vec k_{J,1}d^2
\vec k_{J,2}}\;,
\label{ff}
\end{equation}
where the $i,j$ indices specify parton types (quarks $q=u,d,s$; antiquarks
$\bar q=\bar u,\bar d, \bar s$; or gluon $g$), $f_i(x,\mu)$ denotes
the initial proton parton density function (PDF), the longitudinal
fractions of the partons involved in the hard subprocess are $x_{1,2}$,
$\mu$ is the factorization scale and $d\hat \sigma_{i,j}(x_1 x_2 s,\mu)$ is the
partonic cross section for the production of jets, and $\hat s=x_1 x_2 s$ is
the energy of parton-parton collision.
At lowest order each jet is represented by a single parton having high
transverse momentum, and the partonic subprocess is given by an elementary
two-to-two scattering. In the discussed Mueller-Navelet kinematics the higher
order contributions to the partonic cross section have to be resummed using
BFKL approach. Such resummation at NLA accuracy depends on the details of jet
definition (jet algorithm) and will be specified later.

A convenient starting point for our discussion is the case of inclusive
forward parton-parton scattering, considered in $D=4+2\varepsilon$ dimensions
to regularize the appearing divergences. Due to the optical theorem,
the cross section is related to the imaginary part of the forward
parton-parton scattering amplitude,
\begin{equation}
\hat \sigma =\frac{{\cal I}m_s A}{\hat s} \ ,
\end{equation}
which is given in the BFKL approach at NLA by
\begin{equation}
{\cal I}m_s
 {\cal A} =\frac{\hat s}{(2\pi)^{D-2}}\int\frac{d^{D-2}\vec q_1}{\vec
q_1^{\,\, 2}}\Phi_{P,1}(\vec q_1,s_0)\int
\frac{d^{D-2}\vec q_2}{\vec q_2^{\,\,2}} \Phi_{P,2}(-\vec q_2,s_0)
\int\limits^{\delta +i\infty}_{\delta
-i\infty}\frac{d\omega}{2\pi i}\left(\frac{\hat s}{s_0}\right)^\omega
G_\omega (\vec q_1, \vec q_2)\, ,
\label{PP-scat}
\end{equation}
where the Green's function obeys the BFKL equation
\begin{equation}
\omega \, G_\omega (\vec q_1,\vec q_2)  =\delta^{(D-2)} (\vec q_1-\vec q_2)
+\int d^{D-2}\vec q \, K(\vec q_1,\vec q) \,G_\omega (\vec q, \vec q_1) \;,
\end{equation}
and $\Phi_{P,1}(\vec q_1,s_0)$, ~$\Phi_{P,2}(-\vec q_2,s_0)$
are the parton impact factors calculated
separately for the cases of massless quark and gluon 
in~\cite{fading,fadinq,Cia}.
Here $\vec q_1$, $\vec q_2$ are the  transverse momenta of the Reggeized
gluons, the energy scale parameter $s_0$ is arbitrary, the amplitude, indeed,
does not depend on its choice within NLA accuracy.

\subsection{Parton impact factors}

In this subsection we review the definition of the impact factors in NLO.

Both the kernel of the equation for the Green's function and the parton impact
factors can be expressed in terms of the gluon Regge trajectory,
\begin{equation}
j(t)\;=\;1\:+\:\omega (t)\;,
\end{equation}
and the effective vertices for the Reggeon-parton interaction.

\begin{figure}[tb]
\centering
\includegraphics[]{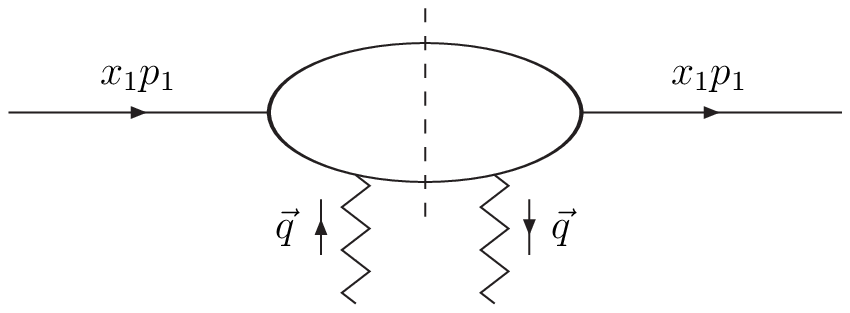}
\caption{Schematic representation of the forward parton impact factor. Here
$p_1$ is the proton momentum, $x_1$ is the fraction of proton momentum
carried by the parton and $\vec q$ is the transverse momentum of the
incoming Reggeized gluon.}
\label{fig:if}
\end{figure}

To be more specific, we will give below the formulae for the case of forward
quark impact factor.
We start with the LO, where the quark impact factors are given by
\begin{equation}
\Phi _{q}^{(0)}(\vec q\,)\;=\;\sum_{\{a\}}\:\int \:\frac{dM_a^2}{2\pi }
\:\Gamma_{a q}^{(0)}(\vec q\,)\:[\Gamma_{a q}^{(0)}(\vec q\,)]^*
\:d\rho_a\;,
\label{eq:a7}
\end{equation}
where $\vec q$ is the Reggeon transverse momentum, and $\Gamma ^{(0)}_{a q}$
denotes the Reggeon-quark vertices in the LO or Born approximation.
The sum $\{a\}$ is over all intermediate states $a$ which contribute to the
$q\rightarrow q$ transition. The phase space element $d\rho_a$ of a state $a$,
consisting of particles with momenta $\ell_n$, is ($p_q$ is initial quark
momentum)
\begin{equation}
d\rho _a\;=\;(2\pi )^D\:\delta^{(D)} \left( p_q+q-\sum_{n\in a}\ell
_n\right) \:\prod_{n\in a}\:\frac{d^{D-1}\ell _n}{(2\pi )^{D-1}2E_n}\;,
\label{eq:a8}
\end{equation}
while the remaining integration in (\ref{eq:a7}) is over the squared
invariant mass of the state $a$,
\[
M_a^2\;=\;(p_q+q)^2\; .
\]

In the LO the only intermediate state which contributes is a one-quark state,
$\{a\}=q$. The integration in Eq.~(\ref{eq:a7}) with the known Reggeon-quark
vertices $\Gamma _{q q}^{(0)}$ is trivial and the quark impact factor reads
\begin{equation}
\Phi _{q}^{(0)}(\vec q\,
)\;=g^2 \frac{\sqrt{N_c^2-1}}{2N_c} \;,
\label{eq:a77}
\end{equation}
where $g$ is QCD coupling, $\alpha_s=g^2/(4\pi)$, $N_c=3$ is the number of QCD
colors.

In the NLO the expression~(\ref{eq:a7}) for the quark impact factor has to be
changed in two ways. First one has to take into account the radiative
corrections to the vertices,
$$
\Gamma _{q q}^{(0)}\to \Gamma_{q q}=\Gamma_{q q}^{(0)}+\Gamma _{q q}^{(1)} \;.
$$
Secondly, in the sum over $\{a\}$ in~(\ref{eq:a7}), we have to include more
complicated states which appear in the next order of perturbative theory. For
the quark impact factor this is a state with an additional gluon, $a=q g$.
However, the integral over $M_a^2$ becomes divergent when an extra gluon
appears in the final state. The divergence arises because the gluon may be
emitted not only in the  fragmentation region of initial quark, but also in
the central rapidity region. The contribution of the central region must be
subtracted from the impact factor, since it is to be assigned in the BFKL
approach to the Green's function. Therefore the result for the forward quark
impact factor, see~\cite{fadinq} and Fig.~\ref{fig:if}, reads
\begin{eqnarray}
\label{eq:a19}
\Phi _{q}(\vec q\, ,s_0)
=\left(\frac{ s_0}{\vec q^{\: 2} }\right)^{\omega(-\vec q^{\: 2})}
\:\sum_{\{a\}}\:\int \:\frac{dM_a^2}{2\pi }\:\Gamma
_{aq}(\vec q\,)\:[\Gamma _{aq}(\vec q\,)]^*\:d\rho _a\:\theta(s_\Lambda -M_a^2)
&&
\nonumber \\
-\frac{1}{2}\int d^{D-2} k\frac{\vec q^{\,\, 2}}{\vec k^{\, 2}}
\Phi_{q}^{(0)}(\vec k)\mathcal{K}_r^{(0)}(\vec k,\vec q\,)
\ln\left( \frac{s_{\Lambda}^2}{(\vec k-\vec q\,)^2 s_0}\right) \;,
&&
\end{eqnarray}
where the intermediate parameter $s_\Lambda $ should go to infinity. The
second term in the r.h.s. of~(\ref{eq:a19}) is the subtraction
of the gluon emission in the central rapidity region. The dependence on
$s_\Lambda$ vanishes because of the cancellation between the first and second
terms.  $K_r^{(0)}$ is the part of LO BFKL kernel related to real gluon
production
\begin{equation}
\label{eq:a20}
K_r^{(0)}(\vec k,\vec q\,)\;=\;\frac{2 g^2N_c}{(2\pi )^{D-1}}
\frac{1}{(\vec k-\vec q\,)^2} \; .
\end{equation}
The factor in~(\ref{eq:a19}) which involves the Regge trajectory arises
from the change of energy scale~($\vec q^{\: 2}\to s_0$) in the vertices
$\Gamma$. The trajectory function $\omega (t)$ can  be taken here in the
one-loop approximation ($t=-\vec q^{\,\,2}$),
\begin{equation}
\omega (t)\;=\;\frac{g^2t}{(2\pi )^{D-1}}\frac{N_c}2\int \frac{d^{D-2}k}
{\vec k^2(\vec q-\vec k)^2}\;=\;-\;g^2N_c\frac{\Gamma (1-\varepsilon )}
{(4\pi )^{D/2}} \frac{\Gamma^2(\varepsilon )}{\Gamma (2\varepsilon )}
(\vec q^{\,\,2})^\varepsilon \; . \label{eq:b20}
\end{equation}

In the Eqs.~(\ref{eq:a7}) and (\ref{eq:a19}) we suppress for shortness the
color indices (for the explicit form of the vertices see~\cite{fading,fadinq}).
The gluon impact factor $\Phi_g(\vec q\,)$ is defined similarly. In the gluon
case only the single-gluon intermediate state contributes in the LO, $a=g$,
which results in
\begin{equation}
\Phi _{g}^{(0)}(\vec q\,
)\;=\frac{C_A}{C_F}\Phi _{q}^{(0)}(\vec q\,) \; ,
\label{eq:a77a}
\end{equation}
here $C_A=N_c$ and $C_F=(N^2_c-1)/(2N_c)$. Whereas in NLO additional two-gluon,
$a=g g$, and quark-antiquark, $a=q \bar q$, intermediate states have to be
taken into account in the calculation of the gluon impact factor.

\subsection{Jet impact factor}

Similarly to the parton-parton scattering~(\ref{PP-scat}) one can represent
the resummed jet cross section in the form
\begin{equation}
\frac{d\sigma}{dJ_1 dJ_2} =\frac{1}{(2\pi)^{D-2}}\!\int\frac{d^{D-2}\vec q_1}
{\vec q_1^{\,\, 2}}\frac{d\Phi_{J,1}(\vec q_1,s_0)}{dJ_1}\!\int
\frac{d^{D-2}\vec q_2}{\vec q_2^{\,\,2}}\frac{d \Phi_{J,2}(-\vec q_2,s_0)}
{dJ_2}\!\!\int\limits^{\delta +i\infty}_{\delta
-i\infty}\frac{d\omega}{2\pi i}\left(\frac{\hat s}{s_0}\right)^\omega
\!G_\omega (\vec q_1, \vec q_2)\, ,
\label{crX}
\end{equation}
where we introduce jet impact factors differential with respect to the
variables parameterizing the jet phase space,
$$
dJ_1\equiv dx_{J,1}d^{D-2}k_{J,1} \; ,\quad\quad
dJ_2\equiv dx_{J,2}d^{D-2}k_{J,2} \; .
$$

Following~\cite{bar1} we consider our process in the frame of a generic and
infrared safe jet algorithm. In practice, this is done by introducing into the
integration over the partonic phase space a suitably defined function which
identifies the jet momentum with the momentum of one parton or with the sum of
the two or more parton momenta when the jet is originated from the
a multi-parton intermediate state.
In our accuracy the jet can be formed by one parton in LO and by one or two
partons when the process is considered in NLO.
In the simplest case, the jet momentum is identified with the momentum of the
parton in the intermediate state $k$ by the following jet
function~\cite{ellis}:
\begin{equation}
\label{jetF0}
S_J^{(2)}(\vec k;x)=\delta(x-x_J)\delta^{(D-2)}(\vec k-\vec k_J)
\;.
\end{equation}
In the more complicated case when the jet originates from a state of two
partons with momenta $k_1$ and $k_2$, we need another function $S_J^{(3)}$,
whose explicit form is specific for the chosen jet algorithm.
An example of jet selection function in the case of the {\em cone
algorithm} is the following~\cite{ellis}:
\begin{eqnarray}
\label{S_3}
S_J^{(3,\rm{cone})}(\vec k_2,\vec k_1, x\beta_1;x) &=&
S_J^{(2)}(\vec k_2; x(1-\beta_1))\Theta\left(\left[ \Delta y^2
+\Delta\phi^2 \right] -\left[ \frac{|\vec k_1|+|\vec k_2|}{\max(|\vec k_1|,
|\vec k_2|)} R_{\mbox{\scriptsize{cone}}} \right]^2 \right) \nonumber \\
&+& S_J^{(2)}(\vec k_1; x\beta_1)\Theta\left( \left[\Delta y^2+\Delta\phi^2
\right] -\left[ \frac{|\vec k_1|+|\vec k_2|}{ \max(|\vec k_1|,|\vec k_2|)}
R_{\mbox{\scriptsize{cone}}} \right]^2 \right) \\
&+& S_J^{(2)}(\vec k_1+\vec k_2; x)\Theta\left( \left[
\frac{|\vec k_1|+|\vec k_2|}
{\max(|\vec k_1|,|\vec k_2|)} R_{\mbox{\scriptsize{cone}}} \right]^2 -
\left[ \Delta y^2+\Delta\phi^2 \right] \right) \;, \nonumber
\end{eqnarray}
where the Sudakov decomposition of the parton momenta
\begin{eqnarray}
k_1&=&x\beta_1p_1+\frac{\vec k_1^{\, 2}}{x \beta_1 s}p_2+k_{1_\perp}\;,
\;\;\;\;\; k_1^2=0 \;, \\
k_2&=&x\beta_2p_1+\frac{\vec k_2^{\, 2}}{x\beta_2 s}p_2+k_{2_\perp}\;,
\;\;\;\;\; k_2^2=0 \;,
%k_J&=&\beta_Jp_1+\frac{\vec k_J^{\, 2}}{xs\beta_J}p_2+k_{J_\perp}\;,
%\;\;\;\;\; k_J^2=0 \;,
\end{eqnarray}
is used, with $\beta_1+\beta_2=1$ and $\vec k_1+\vec k_2=\vec q$, owing to
momentum conservation in the partonic subprocess.
$R_{\mbox{\scriptsize{cone}}}$ in (\ref{S_3}) is the cone-size parameter,
$\Delta y$ and $\Delta \phi$ are the difference of rapidity and azimuthal
angle in the two parton state, respectively:
\begin{equation}
\Delta y=\ln\left(\frac{1-\beta_1}{\beta_1}\frac{|\vec k_1|}
{|\vec k_2|}\right)\;,
\;\;\;\;\;
\Delta\phi=\arccos\frac{\vec k_1\cdot\vec k_2}{\sqrt{\vec k_1^{\, 2}\vec
k_2^{\, 2}}}\;.
\end{equation}
The three terms in $S_J^{(3,{\rm cone})}$ represent the case in which the jet
is formed by the parton $k_2$ or the parton $k_1$ or both, respectively.

In the generic case, the following relations for the jet function must be
fulfilled in order the jet algorithm be infrared safe:
\begin{eqnarray}
S_J^{(3)}(\vec k_2,\vec k_1, x\beta_1;x) &\longrightarrow &
S_J^{(2)}(\vec k_2;x)\;,
\hspace{2.1cm}\vec k_1\rightarrow 0 \;, \;\;\;\beta_1\rightarrow 0 \;,
\nonumber \\
S_J^{(3)}(\vec k_2,\vec k_1, x\beta_1;x)&\longrightarrow &
S_J^{(2)}(\vec k_1+\vec k_2;x)\;,\;\;\;\;\;\;\;\;\;\;\vec k_1\beta_2\rightarrow
\vec k_2 \beta_1\;, \nonumber \\
S_J^{(3)}(\vec k_2,\vec k_1, x\beta_1;x)&\longrightarrow &
S_J^{(2)}(\vec k_2;x(1-\beta_1))\;, \;\;\;\;\;\vec k_1\rightarrow 0 \;, \\
S_J^{(3)}(\vec k_2,\vec k_1, x\beta_1;x)&\longrightarrow &
S_J^{(2)}(\vec k_1; x\beta_1)\;, \hspace{1.7cm}\vec k_2
\rightarrow 0\;. \nonumber
\label{properties}
\end{eqnarray}
Such reduction of $S^{(3)}_J\to S^{(2)}_J$  is required in order that the
singular contributions generated by the real emission be proportional to the
lowest order cross section. These contributions are canceled with the soft and
collinear singularities arising from the virtual corrections and
the collinear counterterms coming from the PDFs renormalization.

Besides, we assume that the jet selection function $S_J^{(3)}$ is symmetric
under the exchange of the final state parton kinematic variables,
$\beta_1\leftrightarrow \beta_2$ and $\vec k_1\leftrightarrow \vec k_2$,
\begin{equation}
S_J^{(3)}(\vec k_2,\vec k_1, x\beta_1;x)=S_J^{(3)}(\vec k_1,\vec k_2,
x\beta_2;x) \; .
\label{symmS3}
\end{equation}

The collinear counterterms appear due to the replacement of the bare PDFs by
the renormalized physical quantities which obey DGLAP evolution equations, in
the $\overline{\rm MS}$ factorization scheme:
\begin{eqnarray}
&
f_q(x)=f_q(x,\mu_F)-\frac{\alpha_s}{2\pi}\left(\frac{1}{\hat \varepsilon}
+\ln\frac{\mu_F^2}{\mu^2}\right)
\int\limits^1_x\frac{dz}{z}\left[P_{qq}(z)f_q(\frac{x}{z},\mu_F)
+P_{qg}(z)f_g(\frac{x}{z},\mu_F)\right] \;,
&\nonumber\\
&
f_g(x)=f_g(x,\mu_F)-\frac{\alpha_s}{2\pi}\left(\frac{1}{\hat \varepsilon}
+\ln\frac{\mu_F^2}{\mu^2}\right)
\int\limits^1_x\frac{dz}{z}\left[P_{gq}(z)f_q(\frac{x}{z},\mu_F)
+P_{gg}(z)f_g(\frac{x}{z},\mu_F)\right]\;,
\label{DGLAPpdfs}
\end{eqnarray}
where $\frac{1}{\hat \varepsilon}=\frac{1}{\varepsilon}+\gamma_E-\ln (4\pi)
\approx \frac{\Gamma(1-\varepsilon)}{\varepsilon (4\pi)^\varepsilon}$ and the
DGLAP splitting functions are:
\begin{eqnarray}
\label{Pqq}
P_{qq}(z)&=&C_F\left( \frac{1+z^2}{1-z}\right)_+
=C_F\left[\frac{1+z^2}{(1-z)_+}+\frac{3}{2}
\delta(1-z)\right]\;, \\
\label{Pqg}
P_{qg}(z)&=&T_R\left[ z^2+(1-z)^2\right]\;, \;\;\;
\mbox{with} \ T_R=\frac{1}{2}\;, \\
\label{Pgg}
P_{gg}(z)&=&2C_A\left( \frac{z}{(1-z)_+}+\frac{(1-z)}{z}
+z(1-z)\right)+\frac{(11C_A-4N_FT_R)}{6}\delta(1-z) \;,\\
\label{Pgq}
P_{gq}(z)&=&C_F\frac{\left[ 1+(1-z)^2\right]}{z}\;;
\end{eqnarray}
here the plus-prescription is defined by
\begin{equation}
\label{plus}
\int_a^1dx\frac{F(x)}{(1-x)_+}=\int_a^1dx\frac{F(x)-F(1)}{1-x}
-\int_0^adx\frac{F(1)}{1-x}\;.
\end{equation}

The other counterterm is related with QCD charge renormalization, in the $\overline{\rm MS}$ scheme:
\begin{equation}
\alpha_s=\alpha_s(\mu_R)\left[1+\frac{\alpha_s(\mu_R)}{4\pi}
\left( \frac{11C_A}{3}-\frac{2N_F}{3} \right)
\left(\frac{1}{\hat \varepsilon}+\ln\frac{\mu_R^2}{\mu^2}\right)\right]\;.
\label{charge-ren}
\end{equation}

Having the results for the lowest order parton impact factors~(\ref{eq:a77})
and~(\ref{eq:a77a}), we get the jet impact factor at the LO level as
\begin{equation}
\frac{d\Phi^{(0)}_J(\vec q\,)}{dJ}=g^2\frac{\sqrt{N_c^2-1}}{2N_c}\int_0^1 dx
\left(\frac{C_A}{C_F}f_g(x)+\sum_{a=q, \bar q} f_{a}(x)\right)S_J^{(2)}
(\vec q;x)\;,
\end{equation}
given as the sum of the gluon and all possible quark and antiquark PDFs
contributions.
Substituting here the bare QCD coupling and bare PDFs by the renormalized ones,
we obtain the following expressions for the counterterms:
\[
\frac{d\Phi_J(\vec q\,)|_{\rm{charge \ c.t.}}}{dJ}
=   \frac{\alpha_s}{2\pi}\left(\frac{1}{\hat \varepsilon}+\ln\frac{\mu_R^2}
{\mu^2}\right)\left( \frac{11C_A}{6}-\frac{N_F}{3} \right)\, \Phi^{(0)}_q
\]
\begin{equation}
\label{charge.count.t}
\times\int_0^1 dx\left(\frac{C_A}{C_F}f_g(x)+\sum_{a=q, \bar q} f_{a}(x)
\right) S_J^{(2)}(\vec q;x)
\end{equation}
for the charge renormalization, and
$$
\frac{d\Phi_J(\vec q\,)|_{\rm{collinear \ c.t.}}}{dJ}
= -  \frac{\alpha_s}{2\pi}\left(\frac{1}{\hat \varepsilon}+\ln\frac{\mu_F^2}
{\mu^2}\right)\Phi^{(0)}_q
\int\limits^1_{0} \, dx \, S_J^{(2)}(\vec q\,;x) \int\limits^1_{x} \frac{dz}{z}
$$
\begin{equation}
\times
\left[ \sum_{a=q,\bar q}\left( P_{qq}(z)f_{a}\left(\frac{x}{z}\right)
+  P_{qg}(z) f_g\left(\frac{x}{ z}\right) \right)
+\frac{C_A}{C_F}\left( P_{gg}(z) f_g\left(\frac{x}{z}\right)
+  P_{gq}(z)\sum_{a=q,\bar q}f_{a}\left(\frac{x}{z}\right)\right)
\right]
\label{c.count.t}
\end{equation}
for the collinear counterterm. The latter can be rewritten in the form
$$
\frac{d\Phi_J(\vec q\,)|_{\rm{collinear \ c.t.}}}{dJ}
= - \frac{\alpha_s}{2\pi}\left(\frac{1}{\hat \varepsilon}+\ln\frac{\mu_F^2}
{\mu^2}\right)\Phi^{(0)}_q
\int\limits^1_{0} \, d\beta \,  \int\limits^1_{0} \,dx\,
S_J^{(2)}(\vec q\,;\beta x)
$$
\begin{equation}
\times\left[ \sum_{a=q,\bar q}\left( P_{qq}(\beta)f_{a}\left(x\right)
+  P_{qg}(\beta) f_g\left(x\right) \right)
+\frac{C_A}{C_F}\left( P_{gg}(\beta) f_g\left(x\right)
+  P_{gq}(\beta)\sum_{a=q,\bar q}f_{a}\left(x\right)\right)
\right]\; .
\label{c.count.t.a}
\end{equation}

Besides, we present the expression for the BFKL counterterm which, in
accordance to the second line of Eq.~(\ref{eq:a19}), provides the subtraction
of the gluon radiation in the central rapidity region:
\begin{eqnarray}
\label{counterterm}
\frac{d\Phi_J(\vec q\,)|_{\rm{BFKL\ c.t.}}}{dJ} &=&-\Phi^{(0)}_q
\frac{C_A \, g^2}{(2\pi)^{D-1}}\int_0^1 dx
\left(\frac{C_A}{C_F}f_g(x)+\sum_{a=q, \bar q} f_{a}(x)\right)
 \nonumber \\
&\times&\int d^{D-2}\vec k \
\frac{\vec q^{\,\, 2}}
{\vec k^{\, 2}(\vec k-\vec q\,)^2} \ln\left(\frac{s_{\Lambda}^2}{s_0\ \vec k^2}
\right)\, S_J^{(2)}(\vec q-\vec k;x) \;.
\end{eqnarray}

Now we have all the necessary ingredients to perform our calculation of the NLO
corrections to the jet impact factor. As a starting point for our consideration
we will use the results of~\cite{fading,fadinq} for the partonic amplitudes
obtained in the calculation of partonic impact factors, introducing there the
appropriate jet functions: $S^{(2)}_J$ for the amplitudes with one-parton state
in the case of one-loop virtual corrections and $S^{(3)}_J$ for the cases with
two partons in the final state (real emission), in order to define the
corresponding contribution to jet cross sections.

For shortness we will present intermediate results for $V$ structures defined
always as
\begin{equation}
\frac{d\Phi^{(1)}_J(\vec q\,)}
{dJ}\, \equiv \, \frac{\alpha_s}{2\pi}\,  \Phi^{(0)}_q \, V (\vec q\,)\; .
\end{equation}
We will consider separately the subprocesses initiated by the quark and the
gluon PDFs and denote
\begin{equation}
V=V_q+V_g \, .
\end{equation}

\section{NLA jet impact factor: the quark contribution}

\begin{figure}[tb]
\centering
\includegraphics{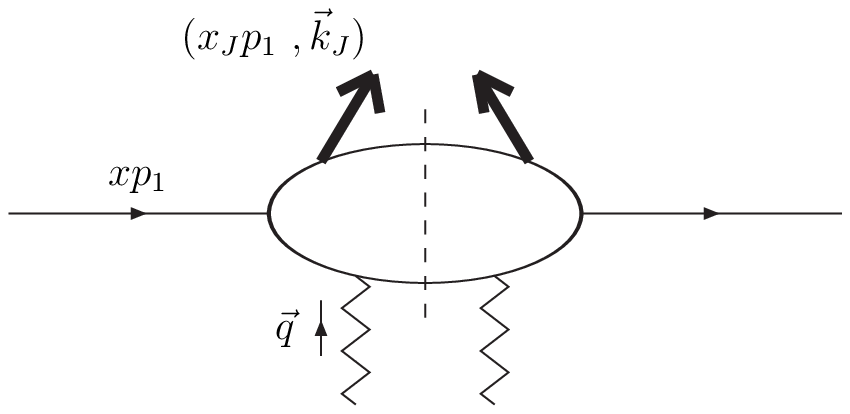}
\caption[]{Schematic representation of the jet vertex for the case of
quark in the initial state. Here $p_1$ is the proton momentum, $x$ is the
fraction of proton momentum carried by the quark, $x_J p_1$ is the
longitudinal jet momentum, $\vec k_J$ is the transverse jet momentum and
$\vec q$ is the transverse momentum of the incoming Reggeized gluon.}
\label{fig:quark}
\end{figure}

We start with the case of incoming quark (see Fig.~\ref{fig:quark}).

\subsection{Virtual correction}

Virtual corrections are the same as in the case of the inclusive quark impact
factor~\cite{fading,fadinq,Cia}:
\[
V_q^{(V)}(\vec q\,)=-\frac{\Gamma[1-\varepsilon]}{\varepsilon\,
(4\pi)^\varepsilon}\frac{\Gamma^2(1+\varepsilon)}{\Gamma(1+2\varepsilon)}
\left(\vec q^{\,\, 2}\right)^{\varepsilon}
\int_0^1 dx \sum_{a=q, \bar q} f_{a}(x) \, S_J^{(2)}(\vec q\,;x)
\]
\[
\times\left\{
C_F\left(\frac{2}{\varepsilon}-\frac{4}{1+2\varepsilon}+1\right)\right.
-N_F\frac{1+\varepsilon}{(1+2\varepsilon)(3+2\varepsilon)}
+C_A\left(\ln\frac{s_0}{\vec q^{\,\,2}}+\psi(1-\varepsilon)-2\psi(\varepsilon)
+\psi(1)\right.
\]
\begin{equation}
\left.\left.
+\frac{1}{4(1+2\varepsilon)(3+2\varepsilon)}-\frac{2}
{\varepsilon(1+2\varepsilon)}
-\frac{7}{4(1+2\varepsilon)}-\frac{1}{2}\right)\right\}\;.
\label{Qvirt}
\end{equation}
Here and in what follows we put the arbitrary scale of dimensional
regularization equal to unity, $\mu=1$. We expand~(\ref{Qvirt}) in
$\varepsilon$ and get
\[
V_q^{(V)}(\vec q\,)=-\frac{\Gamma[1-\varepsilon]}{\varepsilon \,
(4\pi)^\varepsilon}\frac{\Gamma^2(1+\varepsilon)}{\Gamma(1+2\varepsilon)}
\left(\vec q^{\,\, 2}\right)^{\varepsilon} \int_0^1 dx \sum_{a=q, \bar q}
f_{a}(x) \, S_J^{(2)}(\vec q\,;x)
\]
\[
\times \left[C_F\left(\frac{2}{\varepsilon}-3\right)-\frac{N_F}{3}
+C_A\left(\ln\frac{s_0}{\vec q^{\,\,2}}+\frac{11}{6}\right)
\right.
\]
\begin{equation}
\left.
+ \, \varepsilon \left\{8 \,C_F+\frac{5N_F}{9}
-C_A\left(\frac{85}{18}+\frac{\pi^2}{2}\right)\right\}
 \right] +  {\cal O}(\varepsilon) \, .
\label{Qvirt-ex}
\end{equation}

\subsection{Real corrections}

For the incoming quark case, real corrections originate from the quark-gluon
intermediate state. We denote the momentum of the gluon by $k$, then the
momentum of the quark is $q-k$; the longitudinal fraction of the gluon
momentum is denoted by $\beta x$. Thus, the real contribution has the 
form~\cite{fading,fadinq,Cia,Ciafaloni:1998kx}
\begin{eqnarray}
\label{real}
V^{(R)}_q(\vec q\,)&=&\frac{1}
{(4\pi)^{\varepsilon}}\int_0^1dx \sum_{a=q, \bar q} f_{a}(x)
\int\frac{ d^{D-2}\vec k}{\pi^{1+\varepsilon}}
\int_{\beta_0}^{1} d\beta \, {\cal P}_{gq}(\varepsilon,\beta) \nonumber \\
&\times&\frac{\vec q^{\: 2}}{\vec k^{\, 2}(\vec q-\vec k)^2
(\vec k-\beta\vec q\,)^2}
\left\lbrace C_F\beta^2(\vec q-\vec k)^2+C_A(1-\beta)\vec k
\cdot(\vec k-\beta\vec q\,)\right\rbrace \nonumber \\
&\times& S_J^{(3)}\left(\vec q-\vec k, \vec k, x\beta; x\right)\;,
\end{eqnarray}
where
\[
\beta_0=\frac{\vec k^{\, 2}}{s_{\Lambda}}\;, \;\;\;\;\;\;\;
{\cal P}_{gq}(\varepsilon,\beta)=\frac{1+(1-\beta)^2+\varepsilon\beta^2}
{\beta}\;.
\]
The low limit in the $\beta$-integration appears due to the restriction on the
invariant mass of intermediate state, which enters the
definition~(\ref{eq:a19}) of NLO impact factor. Since
\[
M^2_{qg}=\frac{\vec k^2}{\beta}+\frac{(\vec q-\vec k)^2}{1-\beta}-\vec q^{\,\,
2}\;,\;\;\;\; M^2_{qg}\leq s_\Lambda
\; ,
\]
and assuming $s_\Lambda\to \infty$, we obtain that $\beta\geq\beta_0$.

We consider separately the term proportional to $C_F$ and to $C_A$.

The $C_F$-term is not singular for $\beta \to 0$, therefore the limit
$s_\Lambda \to \infty$, or $\beta_0 \to 0$, can be safely taken. We get
\begin{eqnarray}
\label{realCF}
V^{(R)(C_F)}_{q}(\vec q\,)&=&\frac{C_F}
{(4\pi)^{\varepsilon}}\int_0^1dx \sum_{a=q, \bar q} f_{a}(x)\int
\frac{ d^{D-2}\vec k}{\pi^{1+\varepsilon}}
\int_{0}^{1} d\beta \, {\cal P}_{gq}(\varepsilon,\beta) \nonumber \\
&\times&\frac{\vec q^{\: 2} \beta^2}{\vec k^{\, 2}(\vec k-\beta \vec q\,)^2}
  S_J^{(3)}\left(\vec q-\vec k, \vec k, x\beta; x\right)\;.
\end{eqnarray}
In order to isolate all divergences, it is convenient to perform the
change of variable $\vec k=\beta\vec l$ and to present the integral in the form
\begin{eqnarray}
V^{(R)(C_F)}_{q}(\vec q\,)&=&\frac{C_F}
{(4\pi)^{\varepsilon}}\int_0^1dx \sum_{a=q, \bar q} f_{a}(x)
\int_{0}^{1} d\beta \, {\cal P}_{gq}(\varepsilon,\beta) \beta^{2\varepsilon} \\
&\times& \int\frac{ d^{D-2}\vec l}{\pi^{1+\varepsilon}}
\frac{\vec q^{\: 2}}{\vec l^{\: 2}+(\vec l-\vec q\,)^2}
\left[ \frac{1}{(\vec l-\vec q\,)^2}+\frac{1}{\vec l^{\: 2}} \right]
  S_J^{(3)}\left(\vec q-\beta \vec l, \beta \vec l, x\beta; x\right)\;.
\nonumber
\end{eqnarray}
The soft divergence appears for $\beta\rightarrow 0$; in this region
we can introduce the counterterm
\begin{eqnarray}
V^{(R)(C_F, {\rm soft})}_{q}(\vec q\,)&=&\frac{C_F}
{(4\pi)^{\varepsilon}}\int_0^1dx \sum_{a=q, \bar q} f_{a}(x)
\int_{0}^{1} d\beta \, \frac{2}{ \beta^{1-2\varepsilon}} \\
&\times& \int\frac{ d^{D-2}\vec l}{\pi^{1+\varepsilon}}\frac{\vec q^{\: 2}}
{\vec l^{\: 2}+(\vec l-\vec q\,)^2}
\left[ \frac{1}{(\vec l-\vec q\,)^2}+\frac{1}{\vec l^{\: 2}} \right]
  S_J^{(2)}\left(\vec q; x\right)\;,\nonumber
\end{eqnarray}
which equals
\begin{equation}
V^{(R)(C_F, {\rm soft})}_{q}(\vec q\,)=\frac{2 C_F}{\varepsilon}
\frac{\Gamma[1-\varepsilon]}{\varepsilon \,(4\pi)^\varepsilon}
\frac{\Gamma^2(1+\varepsilon)}{\Gamma(1+2\varepsilon)}\left(\vec q^{\,\, 2}
\right)^{\varepsilon}\int_0^1 dx \sum_{a=q, \bar q} f_{a}(x) \,
S_J^{(2)}(\vec q\,;x)\; .
\label{Qreal-soft}
\end{equation}
Collinear divergences arise for $\vec l-\vec q=0$ and for $\vec l=0$;
in these regions we can isolate the two following counterterms:
\begin{eqnarray}
\label{coll1}
V^{(R)(C_F, {\rm coll_1})}_q(\vec q\,)&=&\frac{C_F}
{(4\pi)^{\varepsilon}}\int \frac{d^{D-2}\vec l}{\pi^{1+\varepsilon}
(\vec l-\vec q\,)^2} \Theta (\Lambda^2-(\vec l-\vec q\,)^2) \\
\nonumber
&\times& \int_0^1 d\beta \beta^{2\varepsilon}
\left[{\cal P}_{gq}(\varepsilon,\beta)
-\frac{2}{\beta}\right] \int_0^1dx \sum_{a=q, \bar q} f_{a}(x)
S_J^{(2)}(\vec q\,;x)\;,
\end{eqnarray}
\begin{eqnarray}
\label{coll(2)}
V^{(R)(C_F, {\rm coll_2})}_q(\vec q\,)&=&\frac{C_F}
{(4\pi)^{\varepsilon}}\int \frac{d^{D-2}\vec l}{\pi^{1+\varepsilon}
\vec l^{\: 2}}
\Theta(\Lambda^2 -\vec l^{\: 2}) \int_0^1dx \sum_{a=q, \bar q} f_{a}(x) \\
&\times& \int_0^1 d\beta\
\beta^{2\varepsilon} \left[ S_J^{(2)}(\vec q\,;x(1-\beta))
{\cal P}_{gq}(\varepsilon,\beta)-\frac{2}{\beta}S_J^{(2)}(\vec q\,;x)
\right]\;. \nonumber
\end{eqnarray}
In both these expressions we have introduced an arbitrary cutoff parameter
$\Lambda$ and subtracted the soft divergence. After a simple calculation we
obtain
\begin{eqnarray}
\label{coll1a}
V^{(R)(C_F, {\rm coll_1})}_q(\vec q\,)&=&
\frac{\Gamma[1-\varepsilon]}{\varepsilon \,(4\pi)^\varepsilon}
\frac{\Gamma^2(1+\varepsilon)}{\Gamma(1+2\varepsilon)}
\left(\Lambda^{2}\right)^{\varepsilon}
 \int_0^1dx \sum_{a=q, \bar q} f_{a}(x) S_J^{(2)}(\vec q\,;x) \nonumber \\
&\times& C_F \left[-\frac{3}{2}+4\varepsilon \right]+\mathcal{O}(\varepsilon)
\;.
\end{eqnarray}
The term $V^{(R)(C_F, {\rm coll_2})}_q$ can be rewritten in the following form:
\begin{eqnarray}
\label{coll2a}
V^{(R)(C_F, {\rm coll_2})}(\vec q\,)&=&
\frac{\Gamma[1-\varepsilon]}{\varepsilon \,(4\pi)^\varepsilon}
\frac{\Gamma^2(1+\varepsilon)}{\Gamma(1+2\varepsilon)}
\left(\Lambda^{2}\right)^{\varepsilon}
 \int_0^1 dx \sum_{a=q, \bar q} f_{a}(x)
\left\lbrace -\frac{3}{2}C_FS_J^{(2)}(\vec q\,;x) \right.
\nonumber \\
\nonumber
&+&\int_0^1 d\beta\left[ P_{qq}(\beta)+\, 2\varepsilon (1+\beta^2)
\left(\frac{\ln(1-\beta)}{1-\beta}\right)_+C_F + \varepsilon\,
C_F(1-\beta)\right]
\\
&\times&\left. S_J^{(2)}(\vec q\,; x\beta)\right\rbrace
+\mathcal{O}(\varepsilon)\;,
\end{eqnarray}
where we performed the change of variable $\beta\to 1-\beta$,
used the plus-prescription~(\ref{plus}) and the expansion
\[
(1-\beta)^{2\varepsilon-1}=\frac{1}{2\varepsilon}\delta(1-\beta)
+\frac{1}{(1-\beta)_+}
+2\varepsilon\left( \frac{\ln(1-\beta)}{1-\beta}\right)_+
+\mathcal{O}(\varepsilon^2)\;.
\]
Finally, we can define the term
\begin{equation}
\label{finiteC_F}
V^{(R)(C_F, {\rm finite})}_q=V^{(R)(C_F)}_q-V^{(R)(C_F, {\rm soft})}_q
-V^{(R)(C_F, {\rm coll_1})}_q-V^{(R)(C_F, {\rm coll_2})}_q\;,
\end{equation}
which can be calculated at $\varepsilon=0$.
We remark that $V^{(R)(C_F, {\rm finite})}_q$ and
$V^{(R)(C_F, {\rm cool_{1,2}})}_q$ depend on the cutoff $\Lambda$, but
in the total expression $V^{(R)(C_F)}_q$ this dependence disappears.

The part proportional to $C_A$ in the r.h.s. of Eq.~(\ref{real}) reads
\begin{eqnarray}
V_q^{(R)(C_A)}(\vec q\,)&=&\frac{1}
{(4\pi)^{\varepsilon}}\int_0^1dx \sum_{a=q, \bar q} f_{a}(x)
\int\frac{ d^{D-2}\vec k}{\pi^{1+\varepsilon}}
\int_{\beta_0}^{1} d\beta \, {\cal P}_{gq}(\varepsilon,\beta) \nonumber \\
&\times&\vec q^{\: 2}C_A \,\frac{(1-\beta)\vec k
\cdot(\vec k-\beta\vec q\,) }{\vec k^{\, 2}(\vec q-\vec k)^2
(\vec k-\beta\vec q\,)^2}
 S_J^{(3)}\left(\vec q-\vec k, \vec k, x\beta; x\right)\;.
\end{eqnarray}
The collinear singularity appears at $\vec k-\vec q\,\to 0$; in this
region we can introduce the counterterm
\begin{eqnarray}
\label{collC_A}
V_q^{(R)(C_A, {\rm coll})}(\vec q\,)&=&\frac{C_A}
{(4\pi)^{\varepsilon}} \int\frac{d^{D-2}\vec k}
{\pi^{1+\varepsilon} (\vec q-\vec k)^2}
\Theta\left(\Lambda^2-(\vec k-\vec q\,)^2\right) \nonumber \\
\nonumber
&\times& \int_0^1 dx \sum_{a=q, \bar q} f_{a}(x) \int_0^1\, d\beta \,
{\cal P}_{gq}(\varepsilon,\beta) \, S_J^{(2)}(\vec q\,;x\beta)\\
&=& \frac{\Gamma[1-\varepsilon]}{\varepsilon \,(4\pi)^\varepsilon}
\frac{\Gamma^2(1+\varepsilon)}{\Gamma(1+2\varepsilon)}
\left(\Lambda^{2}\right)^{\varepsilon}  \\
\nonumber
&\times& \int_0^1 dx \sum_{a=q, \bar q} f_{a}(x)  \int_0^1\,  d\beta \,
\left[\frac{C_A}{C_F}P_{gq}(\beta)+ \, \varepsilon \, C_A \, \beta \, \right]
\, S_J^{(2)}(\vec q\,;x\beta) +\mathcal{O}(\varepsilon) \;,
\end{eqnarray}
where $\beta_0$ has been set equal to zero since the expression is finite in
the $\beta\rightarrow0$ limit and, again, the cutoff $\Lambda$ was introduced.
Another singularity appears when $\beta\rightarrow 0$, actually at any value
of gluon transverse momentum $\vec k$. In this region
$S_J^{(3)}\left(\vec q-\vec k, \vec k, x\beta; x\right)\to
S_J^{(2)}\left(\vec q-\vec k; x\right)$ and it is convenient to introduce the
counterterm
\begin{eqnarray}
V_q^{(R)(C_A, {\rm soft} )}(\vec q\,)&=&\frac{C_A}
{(4\pi)^{\varepsilon}}\int_0^1dx \sum_{a=q, \bar q} f_{a}(x)
\int\frac{ d^{D-2}\vec k}{\pi^{1+\varepsilon}}
\int_{\beta_0}^{1} d\beta \,  \frac{2}{\beta}\nonumber \\
 &\times&\vec q^{\: 2} \,\frac{(1-\beta)\vec k
\cdot(\vec k-\beta\vec q\,) }{\vec k^{\, 2}(\vec q-\vec k)^2
(\vec k-\beta\vec q\,)^2}
 S_J^{(2)}\left(\vec q-\vec k; x\right)\nonumber \\
 &=&\frac{C_A}
{(4\pi)^{\varepsilon}}\int_0^1dx \sum_{a=q, \bar q} f_{a}(x)
\int\frac{ d^{D-2}\vec k}{\pi^{1+\varepsilon}}
\int_{\beta_0}^{1} d\beta \,  \frac{2}{\beta} \nonumber \\
&\times&  \,\frac{\vec q^{\: 2}\Theta[(1-\beta)|\vec k|-\beta|\vec q-\vec k|]}
{\vec k^{ 2}(\vec q-\vec k)^2} S_J^{(2)}\left(\vec q-\vec k; x\right)\;,
\label{softCA}
\end{eqnarray}
where the averaging over the relative angle between the vectors $\vec k$ and
$\vec q -\vec k$ has been performed. The integration over $\beta$ gives
the following result for the counterterm:
\begin{eqnarray}
V_q^{(R)(C_A, {\rm soft} )}(\vec q\,)&=&\frac{C_A}
{(4\pi)^{\varepsilon}}\int_0^1dx \sum_{a=q, \bar q} f_{a}(x)
\int\frac{ d^{D-2}\vec k}{\pi^{1+\varepsilon}}
\,\frac{\vec q^{\: 2}}{\vec k^{ 2}(\vec q-\vec k)^2}
\nonumber \\
&\times&
\ln\frac{s^2_\Lambda}{\vec k^2 (|\vec k|+|\vec q-\vec k|)^2}\,\,
S_J^{(2)}\left(\vec q-\vec k; x\right)\;.
\label{softCAa}
\end{eqnarray}
The finite part of the real corrections proportional to $C_A$ is therefore
defined by
\begin{eqnarray}
\label{finiteC_A}
V_q^{(R)(C_A, {\rm finite})}=V_q^{(R)(C_A)}-V_q^{(R)(C_A, {\rm coll})}
-V_q^{(R)(C_A, {\rm soft})}\;.
\end{eqnarray}

When the quark part of BFKL counterterm, given in~(\ref{counterterm}),
\begin{equation}
V_q^{(C)}(\vec q\,)=-\frac{C_A}
{(4\pi)^{\varepsilon}} \int_0^1dx \sum_{a=q, \bar q} f_{a}(x)
\int\frac{d^{D-2}\vec k}
{\pi^{1+\varepsilon} } \ln\left(\frac{s_\Lambda^2}{s_0 \vec k^2}\right)
\frac{\vec q^{\: 2}}{\vec k^{\, 2}(\vec q-\vec k)^2}
 S_J^{(2)}\left(\vec q-\vec k\,;\, x\right)\, ,
\end{equation}
is combined
with $V_q^{(R)(C_A, {\rm soft})}$ given in~(\ref{softCAa}), we see that the
dependence on $s_{\Lambda}$ disappears, as expected, and we get
\begin{eqnarray}
\label{softCA+C}
&&V_q^{(R)(C_A, {\rm soft})}(\vec q\,)+V_q^{(C)}(\vec q\,) \\
&&=\frac{ C_A}
{(4\pi)^{\varepsilon}} \int_0^1 dx \sum_{a=q, \bar q} f_{a}(x)
\int \frac{ d^{D-2}\vec k}{\pi^{1+\varepsilon}}
\frac{\vec q^{\: 2}}{\vec k^{\, 2}(\vec k-\vec q\,)^2}
\ln\left(\frac{s_0}{(|\vec k|+|\vec q-\vec k|)^2}\right)
\ S^{(2)}_J(\vec q-\vec k;x)\;. \nonumber
\end{eqnarray}

\subsection{Final result for the quark in the initial state}

We collect first the contributions given
in~(\ref{Qvirt-ex}),~(\ref{Qreal-soft}),~(\ref{coll1a}),~(\ref{coll2a})
and~(\ref{collC_A}):
\begin{eqnarray}
\label{singular}
&&V_q^{(1)}(\vec q\,)\equiv\left(V_q^{(V)}+V_q^{(R)(C_F, {\rm soft})}
+V_q^{(R)(C_F, {\rm coll_1})}
+V_q^{(R)(C_F, {\rm coll_2})}+V_q^{(R)(C_A, {\rm coll})}\right)(\vec q\,)
\nonumber \\
&=& \frac{\Gamma[1-\varepsilon]}{\varepsilon \,(4\pi)^\varepsilon}
\frac{\Gamma^2(1+\varepsilon)}{\Gamma(1+2\varepsilon)}
\int_0^1 dx\, \sum_{a=q, \bar q} f_{a}(x) \\
\nonumber
&\times&\left\lbrace \left[
\left(\vec q^{\,\, 2}\right)^{\varepsilon} \left(\frac{N_F}{3}
-C_A\ln\left(\frac{s_0} {\vec q^{\: 2}}\right) -
\frac{11 C_A}{6}\right)\,+\, \varepsilon\left(C_A\left( \frac{85}{18}
+\frac{\pi^2}{2}\right) -\frac{5}{9}N_F\right.\right.\right.\\
\nonumber
&+&\left.\left.
C_F\left( 3\ln\frac{\vec q^{\: 2}}{\Lambda^2}-4\right)\right)\right]
\ S^{(2)}_J(\vec q\,;x) \\
\nonumber
&+& \left(\Lambda^{2}\right)^{\varepsilon} \int_0^1 d\beta\left[P_{qq}(\beta)
+\frac{C_A}{C_F}P_{gq}(\beta)\right] \ S^{(2)}_J(\vec q\,;x\beta) \\
\nonumber
&+&\left. \varepsilon
\int_0^1 d\beta\left[2(1+\beta^2)\left(\frac{\ln(1-\beta)}{1-\beta}
\right)_+C_F + C_F(1-\beta)
+ C_A\beta\right]\ S^{(2)}_J(\vec q\,;x\beta)\,
\right\rbrace+\mathcal{O}(\varepsilon)\;.
\end{eqnarray}

Then, we collect the finite contributions, given in Eqs.~(\ref{finiteC_F})
and~(\ref{finiteC_A}), transforming them to the form used in~\cite{bar1}:
\[
V_q^{(2)}(\vec q\,)\equiv\left(V_q^{(R)(C_F, {\rm finite})}
+V_q^{(R)(C_A, {\rm finite})}
\right)(\vec q\,)
\]
\[
=
\int_0^1 dx\, \sum_{a=q, \bar q} f_{a}(x)
\left[ C_F \int_0^1 d\beta \frac{1}{(1-\beta)_+}(1+\beta^2)
\int
\frac{d^{2}\vec l}{\pi \, \vec l^{\: 2}}\left[\frac{\vec q^{\,\, 2}}
{\vec l^2+(\vec q-\vec l)^2}
\right.\right.
\]
\[
\left\{S_J^{(3)}(\vec q-(1-\beta)\vec l,(1-\beta)\vec l,x(1-\beta);x)+
S_J^{(3)}(\vec q \beta+(1-\beta)\vec l,(1-\beta)(\vec q-\vec l),x(1-\beta);x)
\right\}
\]
\[
\biggl.
-\Theta(\Lambda^2-\vec l^{\: 2})
\left\{S_J^{(2)}(\vec q\,;x \beta )+S_J^{(2)}(\vec q\,;x)
\right\}\biggr]
\]
\begin{equation}
+ \, C_A \,\int
\frac{d^{2}\vec k}{\pi\vec k^2}\int_0^1 d\beta\left\{\frac{1+(1-\beta)^2}
{\beta}
\right.
\label{vertex}
\end{equation}
\[
\times\left[
\frac{\vec q^{\: 2}(1-\beta)(\vec q-\vec k)
\cdot(\vec q (1-\beta)-\vec k)}
{(\vec q-\vec k)^2(\vec q (1-\beta)-\vec k)^2}
S_J^{(3)}(\vec k,\vec q-\vec k,x\beta;x)
\right.
\]
\[
\left.\left.\left.
-\Theta(\Lambda^2-\vec k^2)
S_J^{(2)}\left(\vec q\,;x\beta\right)\right]
-\frac{2 \, \vec q^{\: 2}\Theta[(1-\beta)|\vec q - \vec k|-\beta |\vec k|]
}
{\beta (\vec q-\vec k)^2}S_J^{(2)}(\vec k;x)
\right\}\right] +\mathcal{O}(\varepsilon)\, .
\]

Besides, we define
\[
V_q^{(3)}(\vec q\,)\equiv\left(V_q^{(R)(C_A, {\rm soft})}+V_q^{(C)}
\right)(\vec q\,) \; ,
\]
given in Eq.~(\ref{softCA+C}).

Another contribution originates from the collinear and charge renormalization
counterterms, see Eqs.~(\ref{charge.count.t}) and~(\ref{c.count.t.a}),
\[
V_q^{(4)}(\vec q\,)=\frac{\Gamma[1-\varepsilon]}{\varepsilon
\,(4\pi)^\varepsilon}
\int_0^1 dx\, \sum_{a=q, \bar q} f_{a}(x)
\left[\left(\mu_R^2\right)^\varepsilon
\left(\frac{11C_A}{6}-\frac{N_F}{3} \right) S_J^{(2)}\left(\vec q\,;x\right)
\right.
\]
\begin{equation}
\left.
-
\left(\mu_F^{2}\right)^{\varepsilon} \int_0^1 d\beta\left[P_{qq}(\beta)
+\frac{C_A}{C_F}P_{gq}(\beta)\right] \ S^{(2)}_J(\vec q\,;x\beta)\right] \; .
\label{c.t.-q}
\end{equation}

Finally, the quark part of the jet impact factor is given by the sum of the
above four contributions and can be presented as the sum of two terms:
\[
V_q^{(I)}(\vec q\,)=
\int_0^1 dx\, \sum_{a=q, \bar q} f_{a}(x)\left[\frac{ C_A}
{(4\pi)^{\varepsilon}} \int \frac{ d^{D-2}\vec k}{\pi^{1+\varepsilon}}
\frac{\vec q^{\: 2}}{\vec k^{\, 2}(\vec k-\vec q\,)^2}
\ln\frac{s_0}{(|\vec k|+|\vec q-\vec k|)^2} \ S^{(2)}_J(\vec k;x)\right.
\]
\begin{equation}
\label{singular2}
\left. -C_A\ln\left(\frac{s_0}
{\vec q^{\: 2}}\right) \left(\vec q^{\,\, 2}\right)^{\varepsilon}
\frac{\Gamma[1-\varepsilon]}{\varepsilon \,(4\pi)^\varepsilon}
\frac{\Gamma^2(1+\varepsilon)}{\Gamma(1+2\varepsilon)}
S^{(2)}_J(\vec q\,;x) \right] \; .
\end{equation}
and
\begin{equation}
V_q^{(II) }(\vec q\,)=V_q^{(2) }(\vec q\,)+
\int_0^1 dx\, \sum_{a=q, \bar q} f_{a}(x)
\label{non-singular2}
\end{equation}
\[
\times \left\lbrace \left[
 \left(\frac{N_F}{3}  -
\frac{11 C_A}{6}\right)\,\ln\frac{\vec q^{\,\, 2}}{\mu_R^2} \,
+\, C_A\left( \frac{85}{18}+\frac{\pi^2}{2}\right) -\frac{5}{9}N_F
\right.\right. +\left.
C_F\left( 3\ln\frac{\vec q^{\: 2}}{\Lambda^2}-4\right)\right]
\ S^{(2)}_J(\vec q\,;x)
\]
\[
+ \int_0^1 d\beta\left[P_{qq}(\beta)
+\frac{C_A}{C_F}P_{gq}(\beta)\right] \,\ln\frac{\Lambda^2}{\mu_F^2}
\, \ S^{(2)}_J(\vec q\,;x\beta)
\]
\[
+\left.
\int_0^1 d\beta\left[2\left(\frac{\ln(1-\beta)}{1-\beta}
\right)_+ (1+\beta^2) \,C_F + C_F(1-\beta)
+ C_A\beta\right]\ S^{(2)}_J(\vec q\,;x\beta)\,
\right\rbrace\;.
\]

\begin{figure}[tb]
\centering
\includegraphics{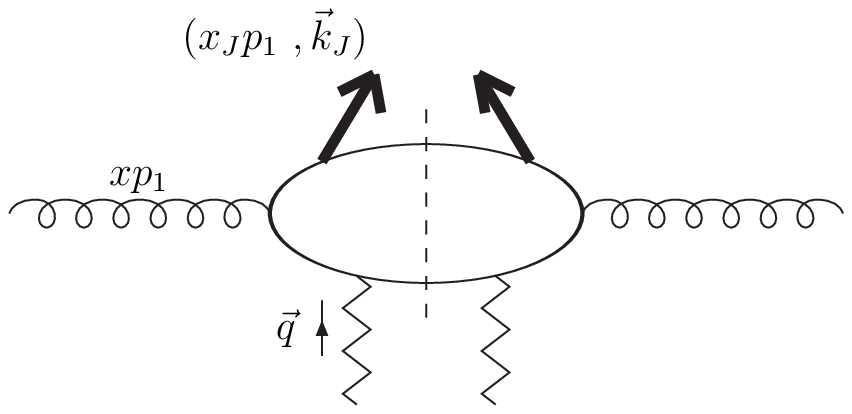}
\caption[]{Schematic representation of the jet vertex for the case of
gluon in the initial state. Here $p_1$ is the proton momentum, $x$ is the
fraction of proton momentum carried by the gluon, $x_J p_1$ is the
longitudinal jet momentum, $\vec k_J$ is the transverse jet momentum and
$\vec q$ is the transverse momentum of the incoming Reggeized gluon.}
\label{fig:gluon}
\end{figure}

\section{NLA jet impact factor: the gluon contribution}

We consider now the case of incoming gluon (see Fig.~\ref{fig:gluon}).

\subsection{Virtual corrections}

Virtual corrections are the same as in the case of the inclusive gluon impact
factor~\cite{fading,fadinq,Cia}:
\begin{eqnarray}
\label{virtualg}
V_g^{(V)}(\vec q)&=&-
\frac{\Gamma[1-\varepsilon]}{\varepsilon\,  (4\pi)^\varepsilon}
\frac{\Gamma^2(1+\varepsilon)}{\Gamma(1+2\varepsilon)}
\left(\vec q^{\,\, 2}\right)^{\varepsilon}\,
\int_0^1 dx \, \frac{C_A}{C_F}\, f_g(x) \, S_J^{(2)}(\vec q\,;x) \\
\nonumber
&\times&\left[C_A\ln\left(\frac{s_0}{\vec q^{\: 2}}\right)
+C_A\left(\frac{2}{\varepsilon}-\frac{11+9\varepsilon}{2(1+2\varepsilon)
(3+2\varepsilon)}\right.\right. \\
\nonumber
&+&\left.\frac{N_F}{C_A}\frac{(1+\varepsilon)(2+\varepsilon)-1}
{(1+\varepsilon)(1+2\varepsilon)(3+2\varepsilon)}+\psi(1)
+\psi(1-\varepsilon)-2\psi(1+\varepsilon)\right) \\
\nonumber
&+&\left. C_A\frac{\varepsilon}{(1+\varepsilon)(1+2\varepsilon)
(3+2\varepsilon)}\left(1+\varepsilon-\frac{N_F}{C_A}\right)
\frac{1}{(1+\varepsilon) }\right] \;.
\end{eqnarray}
The $\varepsilon$-expansion has the form
\begin{eqnarray}
\label{virtualg-e}
V_g^{(V)}(\vec q)&=&-
\frac{\Gamma[1-\varepsilon]}{\varepsilon\,  (4\pi)^\varepsilon}
\frac{\Gamma^2(1+\varepsilon)}{\Gamma(1+2\varepsilon)}
\left(\vec q^{\,\, 2}\right)^{\varepsilon}\,
\int_0^1 dx \, \frac{C_A}{C_F}\, f_g(x) \, S_J^{(2)}(\vec q\,;x) \nonumber\\
\nonumber
&\times&\left[\, C_A\left( \ln\left(\frac{s_0}{\vec q^{\: 2}}\right)
+\frac{2}{\varepsilon}-\frac{11}{6}\right) + \frac{N_F}{3}\right. \\
&+& \left. \varepsilon \left\{ \, C_A\left(\frac{67}{18}
-\frac{\pi^2}{2}\right)-\frac{5}{9}N_F \right\} \right]
+ \mathcal{O}(\varepsilon) \; .
\end{eqnarray}

\subsection{Real corrections: $q \bar q$ intermediate state}

In the NLO gluon impact factor real corrections come from intermediate states
of two particles, which can be quark-antiquark or gluon-gluon. In this
Subsection we consider the former. The real contribution from the
quark-antiquark case is ~\cite{fading,fadinq,Cia,Ciafaloni:1998kx}:
\begin{eqnarray}
\label{real-qq}
V_g^{(R_{q\bar q})}(\vec q\,)&=& \frac{N_F }
{(4\pi)^{\varepsilon}}\int_0^1 dx \, \frac{C_A}{C_F} \, f_g(x)
\int_0^1 d\beta \,\int \frac{d^{D-2}\vec k}{\pi^{1+\varepsilon}}
\, \frac{\vec q^{\,\, 2}}{\vec k^{\, 2}(\vec k-\vec q\,)^2}\\
\nonumber
&\times&T_R \,{\cal P}_{qg}(\varepsilon,\beta)\left[\frac{C_F}{C_A}+
\frac{\beta(1-\beta)\vec k\cdot(\vec q-\vec k)}
{(\vec k-\beta\vec q\,)^2}\right]
S_J^{(3)}(\vec q-\vec k,\vec k,x\beta;x)\;,
\end{eqnarray}
with
\begin{equation}
{\cal P}_{qg}(\varepsilon,\beta)=1-\frac{2\beta(1-\beta)}{1+\varepsilon}
\;.
\end{equation}
Below we discuss separately the first and the second contributions in the
r.h.s. of Eq.~(\ref{real-qq}), which we denote $V_g^{(R_{q\bar q})(C_F)}$
and $V_g^{(R_{q\bar q})(C_A)}$.

The first contribution is
\begin{eqnarray}
V_g^{(R_{q\bar q})(C_F)}(\vec q\,)&=&\frac{N_F}
{(4\pi)^{\varepsilon}}\int_0^1 dx \, \frac{C_A}{C_F} \, f_g(x)\int_0^1 d\beta\,
\int \frac{d^{D-2}\vec k}{\pi^{1+\varepsilon}}
\frac{\vec q^{\: 2}}{\vec k^{\, 2}(\vec k-\vec q\,)^2} \nonumber \\
&\times&T_R \,{\cal P}_{qg}(\varepsilon,\beta)\, \frac{C_F}{C_A}\,
S_J^{(3)}(\vec q-\vec k,\vec k,x\beta;x)\\
\nonumber
&=& \frac{N_F }
{(4\pi)^{\varepsilon}}\int_0^1 dx  \, f_g(x)\int_0^1 d\beta\,T_R\,
{\cal P}_{qg}(\varepsilon,\beta) \int \frac{d^{D-2}\vec k}{\pi^{1+\varepsilon}}
\\
\nonumber
&\times&\frac{\vec q^{\: 2}}{\vec k^{ 2}+(\vec q-\vec k)^2}
\left[ \frac{1}{\vec k^{ 2}}+\frac{1}{(\vec q-\vec k)^2}\right]
S_J^{(3)}(\vec q-\vec k,\vec k,x\beta;x)\;.
\end{eqnarray}
Here we have collinear divergences for $\vec k=0$ and $\vec q-\vec k=0$.
The contribution in these kinematical regions is the same, as can be easily
seen after the changes of variables $\vec k\rightarrow \vec q-\vec k$
and $ \beta\rightarrow 1-\beta$, since ${\cal P}_{qg}(\varepsilon,\beta)
={\cal P}_{qg}(\varepsilon,1-\beta)$ and taking into account the
property~(\ref{symmS3}) that the $S_J^{(3)}$ jet selection function has to
possess. Therefore we can write
\begin{eqnarray}
V_g^{(R_{q\bar q})(C_F)}(\vec q\,)
&=& \frac{2 N_F }
{(4\pi)^{\varepsilon}}\int_0^1 dx  \, f_g(x)\int_0^1 d\beta \,
T_R\, {\cal P}_{qg}(\varepsilon,\beta) \nonumber \\
&\times&\int \frac{d^{D-2}\vec k}{\pi^{1+\varepsilon}\, \vec k^2}\, \,
\frac{\vec q^{\: 2}}{ \vec k^{ 2}+(\vec q -\vec k)^2}\,\,
S_J^{(3)}(\vec k,\vec q-\vec k,x \beta;x)
\end{eqnarray}
and isolate the collinearly divergent part given by
\begin{eqnarray}
\label{collC_Fg}
V_g^{(R_{q\bar q})(C_F, {\rm coll})}(\vec q\,)&=&  \frac{2 N_F }
{(4\pi)^{\varepsilon}}\int_0^1 dx  \, f_g(x)\int_0^1 d\beta \,T_R\,
{\cal P}_{qg}(\varepsilon,\beta)\\
\nonumber
&\times& \int \frac{d^{D-2}\vec k}{\pi^{1+\varepsilon}\, \vec k^2 }\,\,
\Theta(\Lambda^2-\vec k^2) \,\,S_J^{(2)}(\vec q;x\beta) \\
\nonumber
&=& 2 N_F \, \frac{\Gamma[1-\varepsilon]}{\varepsilon\,  (4\pi)^\varepsilon}
\frac{\Gamma^2(1+\varepsilon)}{\Gamma(1+2\varepsilon)}
\left(\Lambda^{2}\right)^{\varepsilon}\,\int_0^1 dx  \, f_g(x) \\
\nonumber
&\times&\int_0^1 d\beta \, \left[ P_{qg}(\beta)
+ \varepsilon\beta(1-\beta)\right] \,S_J^{(2)}(\vec q;x\beta)
+\mathcal{O}(\varepsilon) \;,
\end{eqnarray}
where we have introduced, as before, the cutoff parameter $\Lambda$.
The finite part is therefore defined  by
\begin{eqnarray}
\label{finitegC_F}
V_g^{(R_{q \bar q})(C_F, {\rm finite})}=
V_g^{(R_{q\bar q})(C_F)} - V_g^{(R_{q\bar q})(C_F, {\rm coll})}\;,
\end{eqnarray}
where one can take $\varepsilon\to 0$ limit and get
\[
V_g^{(R_{q \bar q})(C_F, {\rm finite})}
= 2 N_F
\, \int_0^1 dx  \, f_g(x)\int_0^1 d\beta \, P_{qg}(\beta)
\]
\begin{equation}
\times\int \frac{d^{2}\vec k}{\pi\, \vec k^2}\, \left[
\frac{\vec q^{\: 2}}{ \vec k^{ 2}+(\vec q-\vec k)^2}
S_J^{(3)}(\vec k,\vec q-\vec k,x\beta;x)-\Theta(\Lambda^2-\vec k^2)
\,\,S_J^{(2)}(\vec q;x\beta) \right]+\mathcal{O}(\varepsilon) \; .
\label{finitegC_Fa}
\end{equation}

The second contribution in~(\ref{real-qq}) is
\begin{eqnarray}
V_g^{(R_{q \bar q})(C_A)}(\vec q \,)&=& \frac{N_F }
{(4\pi)^{\varepsilon}}\int_0^1 dx \, \frac{C_A}{C_F} \, f_g(x)
\int_0^1 d\beta \,\int \frac{d^{D-2}\vec k}{\pi^{1+\varepsilon}}
\, \frac{\vec q^{\,\, 2}}{\vec k^{\, 2}(\vec q-\vec k)^2}\label{gg.CA} \\
\nonumber
&\times&T_R \,{\cal P}_{qg}(\varepsilon,\beta)
\frac{\beta(1-\beta)\vec k\cdot(\vec q-\vec k)}
{(\vec k-\beta\vec q\,)^2}
S_J^{(3)}(\vec q-\vec k,\vec k,x\beta;x)\;.
\end{eqnarray}
Here the collinear divergence appears for $\vec k-\beta\vec q\,=0$ and the
integral in this region can be identified with
\begin{eqnarray}
\label{collC_Ag}
V_g^{(R_{q\bar q})(C_A, {\rm coll})}(\vec q\,)&=& \frac{N_F }
{(4\pi)^{\varepsilon}}\int_0^1 dx \, \frac{C_A}{C_F} \, f_g(x)
\int_0^1 d\beta\, T_R\, {\cal P}_{qg}(\varepsilon,\beta) \nonumber \\
&\times&\int \frac{d^{D-2}\vec k}{\pi^{1+\varepsilon}(\vec k-\beta\vec q\,)^2}
\Theta(\Lambda^2-(\vec k-\beta\vec q\,)^2)S_J^{(2)}(\vec q\,;x) \\
&=& N_F
 \, \frac{\Gamma[1-\varepsilon]}{\varepsilon\,  (4\pi)^\varepsilon}
\frac{\Gamma^2(1+\varepsilon)}{\Gamma(1+2\varepsilon)}
\left(\Lambda^{2}\right)^{\varepsilon}\,
\left(\frac{1}{3}+\frac{\varepsilon}{6}\right) \nonumber \\
&\times&  \int_0^1 dx \, \frac{C_A}{C_F} \, f_g(x) S_J^{(2)}(\vec q\,;x)
+\mathcal{O}(\varepsilon)\;.\nonumber
\end{eqnarray}
Then, the finite part can be written as
\begin{eqnarray}
\label{finitegC_A}
V_g^{(R_{q\bar q})(C_A, {\rm finite})}=V_g^{(R_{q\bar q})(C_A)}
-V_g^{(R_{q\bar q})(C_A, {\rm coll})}\;.
\end{eqnarray}
After the change of variable $\vec k\rightarrow \vec q-\vec k$
in (\ref{gg.CA}) and (\ref{collC_Ag}) we have
\begin{eqnarray}
V_g^{(R_{q \bar q})(C_A, {\rm finite})}(\vec q \,)&=& N_F
\int_0^1 dx \, \frac{C_A}{C_F} \, f_g(x) \int_0^1 d\beta \, P_{qg}(\beta)
\label{finitegC_Aa}\\
&\times& \int \frac{d^{2}\vec k}{\pi\, (\vec k-(1-\beta)\vec q\,)^2}
\, \left[
\frac{\vec q^{\: 2} \beta(1-\beta)\vec k\cdot(\vec q-\vec k)}
{\vec k^{\, 2}(\vec k-\vec q\,)^2}
S_J^{(3)}(\vec k,\vec q-\vec k,x\beta;x)\right.\nonumber \\
&&\left.
-\Theta(\Lambda^2-(\vec k-(1-\beta)\vec q\,)^2)S_J^{(2)}(\vec q\,;x)
\right]  +\mathcal{O}(\varepsilon) \;. \nonumber
\end{eqnarray}

\subsection{Real corrections: $gg$ intermediate state}

The real contribution from the gluon-gluon case is~\cite{fading,fadinq,Cia,
Ciafaloni:1998kx}:
\begin{eqnarray}
V_g^{(R_{gg})}(\vec q\,)&=& \frac{C_A }
{(4\pi)^{\varepsilon}} \int_0^1 dx \, \frac{C_A}{C_F} \, f_g(x)
\int \frac{d^{D-2}\vec k}{\pi^{1+\varepsilon}}
\int_{\beta_0}^{1-\beta_0} d\beta \frac{\vec q^{\: 2}\ {\cal P}_{gg}(\beta)}
{(\vec k-\beta \vec q\,)^2 \vec k^2 (\vec k-\vec q)^2} \nonumber \\
\nonumber
&\times& \left\lbrace \beta^2(\vec k-\vec q\,)^2 + (1-\beta)^2
\vec k^2-\beta(1-\beta)\vec k\cdot(\vec q-\vec k)\right\rbrace \\
&\times& S_J^{(3)}(\vec q-\vec k,\vec k,x\beta;x) \;.
\end{eqnarray}
where
\[
{\cal P}_{gg}(\beta)= P(\beta)+P(1-\beta)\;, \;\;\;\;\; {\rm with}
\;\;\;\;\; P(\beta)=\left(\frac{1}{\beta}+\frac{\beta}{2}\right)
(1-\beta)\;.
\]
We note that here the lower integration limit in $\beta$ is
$\beta_0=\vec k^{\, 2}/s_{\Lambda}$, whereas the upper limit is $1-\beta_0$.
This comes from the $\Theta$ function in the impact factor
definition~(\ref{eq:a19}), which restricts the radiation of either of the two
gluons into the central region of rapidity.

Using the symmetry of the integrand under the change of variables describing
the two gluons, $\beta \to 1-\beta$ and $\vec k\to \vec q-\vec k $ (thanks to
the symmetry property~(\ref{symmS3}) of the jet function), we get
\begin{eqnarray}
\label{real-gg}
V_g^{(R_{gg})}(\vec q\,)&=& 2\, \frac{C_A }
{(4\pi)^{\varepsilon}} \int_0^1 dx \, \frac{C_A}{C_F} \,f_g(x)
\int \frac{d^{D-2}\vec k}{\pi^{1+\varepsilon}}\int_{\beta_0}^{1-\beta_0}
\, d\beta \, P(\beta) \\
\nonumber
&\times&\frac{\vec q^{\: 2}}{(\vec k-\beta \vec q\,)^2 \vec k^2
(\vec k-\vec q\,)^2} \left\lbrace \beta^2(\vec k-\vec q\,)^2
+(1-\beta)\vec k\cdot(\vec k-\beta\vec q\,)\right\rbrace \\
\nonumber
&\times& S_J^{(3)}(\vec q-\vec k,\vec k,x\beta;x)
\equiv V_g^{(R_{gg})(A)}(\vec q\,) +V_g^{(R_{gg})(B)}(\vec q\,)\;.
\end{eqnarray}
In this form the upper limit of $\beta$ integration can be put to unity.

In $V_g^{(R_{gg})(A)}$ the lower integration limit $\beta_0$ can be put equal
to zero.  Then, after the change of variable $\vec k=\beta\vec l$, we obtain
\begin{eqnarray}
V_g^{(R_{gg})(A)}(\vec q\,)&=&  2\, \frac{C_A }
{(4\pi)^{\varepsilon}}  \int_0^1 dx \, \frac{C_A}{C_F} \, f_g(x)\int_0^1 d\beta
P(\beta) \beta^{2\varepsilon}  \int \frac{d^{D-2}\vec l}{\pi^{1+\varepsilon}}
\nonumber \\
&\times&\frac{\vec q^{\: 2}}{\vec l^{\: 2}(\vec l-\vec q)^2}
S_J^{(3)}(\vec q-\beta\vec l,\beta\vec l,x\beta;x) \\
\nonumber
&=& 2\, \frac{C_A }
{(4\pi)^{\varepsilon}}  \int_0^1 dx \, \frac{C_A}{C_F} \, f_g(x)\int_0^1 d\beta
P(\beta) \beta^{2\varepsilon}  \int \frac{d^{D-2}\vec l}{\pi^{1+\varepsilon}}
 \\
\nonumber
&\times&\frac{\vec q^{\: 2}}{\vec l^{\: 2}+(\vec l-\vec q\,)^2}
\left[ \frac{1}{\vec l^{\: 2}}+\frac{1}{(\vec l-\vec q\,)^2}\right]
S_J^{(3)}(\vec q-\beta\vec l,\beta\vec l,x\beta;x)\;.
\end{eqnarray}
In this expression one has both soft and collinear divergences. The soft
divergence can be isolated in the counterterm
\begin{eqnarray}
\label{softA}
V_g^{(R_{gg})(A, {\rm soft})}(\vec q\,)&=& 2\, \frac{C_A }
{(4\pi)^{\varepsilon}}  \int_0^1 dx \, \frac{C_A}{C_F} \, f_g(x)
\int_0^1 \frac{d\beta}{\beta^{1-2\varepsilon}}
\int \frac{ d^{D-2}\vec l}{\pi^{1+\varepsilon}}\frac{\vec q^{\: 2}}
{\vec l^{\: 2}+(\vec l-\vec q\,)^{2}} \nonumber \\
&\times& \left[\frac{1}{(\vec l-\vec q\,)^{2}}+\frac{1}{\vec l^{\: 2}}\right]
S_J^{(2)}(\vec q\,;x) \;,
\end{eqnarray}
which equals
\begin{eqnarray}
\label{softAa}
V_g^{(R_{gg})(A, {\rm soft})}(\vec q\,)&=& \frac{\Gamma[1-\varepsilon]}
{\varepsilon\,  (4\pi)^\varepsilon}
\frac{\Gamma^2(1+\varepsilon)}{\Gamma(1+2\varepsilon)}
\left(\vec q^{\,\, 2}\right)^{\varepsilon}\, \frac{2\, C_A}{\varepsilon}
 \int_0^1 dx \, \frac{C_A}{C_F} \, f_g(x)\,
S_J^{(2)}(\vec q\,;x) \; .
\end{eqnarray}
After the subtraction of the soft divergence, collinear divergences still
appear for $\vec l=0$ and $\vec l - \vec q\,=0$ and can be isolated by the
following two counterterms:
\begin{eqnarray}
\label{collA1}
V_g^{(R_{gg})(A, {\rm coll_1})}(\vec q\,)&=&2\, \frac{C_A }
{(4\pi)^{\varepsilon}} \int_0^1 dx \, \frac{C_A}{C_F} \, f_g(x)
\int \frac{d^{D-2}\vec l}{\pi^{1+\varepsilon}(\vec q-\vec l)^2}\,
\Theta\left(\Lambda^2-(\vec q-\vec l)^2\right)
\nonumber \\
&\times& \int_0^1 d\beta
\beta^{2\varepsilon} \left( P(\beta)-\frac{1}{\beta}\right)
S_J^{(2)}(\vec q\,;x)
\end{eqnarray}
and
\begin{eqnarray}
\label{collA2}
V_g^{(R_{gg})(A, {\rm coll_2})}(\vec q\,)&=&2\, \frac{C_A }
{(4\pi)^{\varepsilon}} \int_0^1 dx \, \frac{C_A}{C_F} \, f_g(x)
\int \frac{d^{D-2}\vec l}{\pi^{1+\varepsilon}\vec l^{\:2}}\,
\Theta\left(\Lambda^2-\vec l^{\: 2}\right)
 \nonumber \\
&\times& \int_0^1 d\beta\, \beta^{2\varepsilon}
\left( P(\beta)S_J^{(2)}(\vec q\,;x(1-\beta))-\frac{1}{\beta}
S_J^{(2)}(\vec q\,;x)\right) \, .
\end{eqnarray}
These counterterms equal
\begin{eqnarray}
\label{collA1a}
V_g^{(R_{gg})(A, {\rm coll_1})}(\vec q\,)&=& \frac{\Gamma[1-\varepsilon]}
{\varepsilon\,  (4\pi)^\varepsilon}
\frac{\Gamma^2(1+\varepsilon)}{\Gamma(1+2\varepsilon)}
\left(\Lambda^2\right)^{\varepsilon}\,  \int_0^1 dx \, \frac{C_A}{C_F}
\, f_g(x) \, S_J^{(2)}(\vec q\,;x) \nonumber \\
&\times& C_A\, \left(-\frac{11}{6}+\, \varepsilon\, \frac{67}{18}\,\right)
+\mathcal{O}(\varepsilon) \, ,
\end{eqnarray}
\begin{eqnarray}
\label{collA2a}
\nonumber
V_g^{(R_{gg})(A, {\rm coll_2})}(\vec q\,)&=&
\frac{\Gamma[1-\varepsilon]}{\varepsilon\,  (4\pi)^\varepsilon}
\frac{\Gamma^2(1+\varepsilon)}{\Gamma(1+2\varepsilon)}
\left(\Lambda^2\right)^{\varepsilon}\,  \int_0^1 dx \, \frac{C_A}{C_F}
\, f_g(x) \int_0^1 d\beta (1-\beta)P(1-\beta) \\
\nonumber
&\times& 2 \,C_A \left[\frac{1}{(1-\beta)_+}+2\varepsilon
\left(\frac{\ln(1-\beta)}
{1-\beta}\right)_+\,\right] \\
&\times& S_J^{(2)}(\vec q\,;x\beta)+\mathcal{O}(\varepsilon) \;,
\end{eqnarray}
where to obtain  the last equation we made the change of variable
$\beta\to 1-\beta$ and expanded the term $(1-\beta)^{2\varepsilon-1}$.
The finite part is therefore defined by
\begin{equation}
\label{finiteA}
V_g^{(R_{gg})(A, {\rm finite})}=V_g^{(R_{gg})(A)}-V_g^{(R_{gg})(A, {\rm soft})}
-V_g^{(R_{gg})(A, {\rm coll_1})}-V_g^{(R_{gg})(A, {\rm coll_2})}\;.
\end{equation}

The $V_g^{(R_{gg})(B)}$ term, defined in~(\ref{real-gg}), has a collinear
divergence for $\vec k-\vec q\,=0$. It can be isolated in the following
integral:
\begin{eqnarray}
\label{collB}
V_g^{(R_{gg})(B, {\rm coll})}(\vec q\,)&=&
2\, \frac{C_A }
{(4\pi)^{\varepsilon}} \int_0^1 dx \, \frac{C_A}{C_F} \, f_g(x)
\int_0^1 d\beta P(\beta) \nonumber \\
&\times&\int \frac {d^{D-2} \vec k}{\pi^{1+\varepsilon}(\vec k-\vec q\,)^2}
\Theta \left(\Lambda^2-(\vec k-\vec q\,)^2 \right)
S_J^{(2)}(\vec q\,;x\beta)\;,
\\
&=&
 \frac{\Gamma[1-\varepsilon]}{\varepsilon\,  (4\pi)^\varepsilon}
\frac{\Gamma^2(1+\varepsilon)}{\Gamma(1+2\varepsilon)}
\left(\Lambda^2\right)^{\varepsilon}\, \int_0^1 dx \, \frac{C_A}{C_F}
\, f_g(x) \nonumber
\\
\nonumber &\times&
\int_0^1 d\beta \, 2 \, C_A\, P(\beta) \, S_J^{(2)}(\vec q\,;x\beta)
+\mathcal{O}(\varepsilon) \; ,\nonumber
\end{eqnarray}
where $\beta_0$ has been put equal to zero thanks to the property of lowest
order jet function~(\ref{jetF0}).

Another singularity appears when $\beta\rightarrow 0$; in this region we
can isolate the term
\begin{eqnarray}
&&V_g^{(R_{gg})(B, {\rm soft})}(\vec q\,)=
\label{beta_1g} \\
&&2\, \frac{C_A }
{(4\pi)^{\varepsilon}} \int_0^1 dx \, \frac{C_A}{C_F} \, f_g(x)
\int \frac{d^{D-2}\vec k}{\pi^{1+\varepsilon}}
\int_{\beta_0}^1  \frac{d\beta}{\beta}\,
% \frac{\vec q^{\: 2}}{\vec k^{ 2}(\vec k-\vec q\,)^2}
\vec q^{\: 2} \,\frac{(1-\beta)\vec k
\cdot(\vec k-\beta\vec q\,) }{\vec k^{\, 2}(\vec q-\vec k)^2
(\vec k-\beta\vec q\,)^2} \ S^{(2)}_J(\vec q-\vec k;x)
\nonumber \\
&&=2\, \frac{C_A }
{(4\pi)^{\varepsilon}} \int_0^1 dx \, \frac{C_A}{C_F} \, f_g(x)
\int \frac{d^{D-2}\vec k}{\pi^{1+\varepsilon}}
\int_{\beta_0}^1  \frac{d\beta}{\beta}\, \frac{\vec q^{\: 2}
\Theta[(1-\beta)|\vec k|-\beta|\vec q-\vec k|] }{\vec k^{ 2}
(\vec k-\vec q\,)^2} \ S^{(2)}_J(\vec q-\vec k;x) \nonumber \\
&&=\, \frac{C_A }
{(4\pi)^{\varepsilon}} \int_0^1 dx \, \frac{C_A}{C_F} \, f_g(x)
\int \frac{d^{D-2}\vec k}{\pi^{1+\varepsilon}}
\frac{\vec q^{\: 2}}{\vec k^{ 2}(\vec k-\vec q\,)^2}
\ln\frac{s^2_\Lambda}{\vec k^2(|\vec k|+|\vec q-\vec k|)^2}
\ S^{(2)}_J(\vec q-\vec k;x)\nonumber\;.
\end{eqnarray}

The finite part of $V_g^{(R_{gg})(B)}$ is therefore defined by
\begin{eqnarray}
\label{finiteB}
V_g^{(R_{gg})(B, {\rm finite})}=V_g^{(R_{gg})(B)}-V_g^{(R_{gg})(B, {\rm coll})}
-V_g^{(R_{gg})(B, {\rm soft})}\;.
\end{eqnarray}

When the gluon part of BFKL counterterm, given in~(\ref{counterterm}),
is combined with  $V_q^{(R_{gg})(B, {\rm soft})}$, given in~(\ref{beta_1g}),
we see that the dependence on $s_{\Lambda}$ disappears and we obtain
\begin{eqnarray}
&&V_g^{(R_{gg})(B, {\rm soft})}(\vec q\,) + V_g^{(C)}(\vec q\,)
\label{softCA+C_gluon} \\
&&= \frac{C_A }
{(4\pi)^{\varepsilon}} \int_0^1 dx \, \frac{C_A}{C_F} \, f_g(x)
\int \frac{d^{D-2}\vec k}{\pi^{1+\varepsilon}}
\frac{\vec q^{\: 2}}{\vec k^{ 2}(\vec k-\vec q\,)^2}
\ln \frac{s_0}{(|\vec k|+|\vec q-\vec k|)^2} \,\ S^{(2)}_J(\vec q-\vec k;x)\;.
\nonumber
\end{eqnarray}

\subsection{Final result for the gluon in the initial state}

We collect first the contributions which contain singularities given
in~(\ref{virtualg-e}),~(\ref{collC_Fg}),~(\ref{collC_Ag}),~(\ref{softAa}),
(\ref{collA1a}),~(\ref{collA2a}) and~(\ref{collB}) and get
\begin{eqnarray}
\label{singularg}
V_g^{(1)}(\vec q\,)&\equiv&\left(V_g^{(V)}+V_g^{(R_{q\bar q})(C_F, {\rm coll})}
+V_g^{(R_{q\bar q})(C_A, {\rm coll})}+V_g^{(R_{gg})(A, {\rm soft})} \right.\\
&+&\left.V_g^{(R_{gg})(A, {\rm coll_1})}+V_g^{(R_{gg})(A, {\rm coll_2})}
+V_g^{(R_{gg})(B, {\rm coll})}\right)(\vec q\,) \nonumber \\
&=&
\frac{\Gamma[1-\varepsilon]}{\varepsilon\,  (4\pi)^\varepsilon}
\frac{\Gamma^2(1+\varepsilon)}{\Gamma(1+2\varepsilon)}\,
\int_0^1 dx \, \frac{C_A}{C_F}\, f_g(x)
\nonumber \\
&\times&\left\lbrace \left[\left(\vec q^{\: 2}\right)^\varepsilon
\left(
\frac{11 C_A}{6}-\frac{N_F}{3} -C_A\ln\left(\frac{s_0}{\vec q^{\: 2}}\right)
\right) -2\left(\Lambda^2\right)^\varepsilon \left( \frac{11 C_A}{6}-\frac{N_F}
{3}\right)
\right. \right.\nonumber \\
&+&\left.\varepsilon\left(\frac{\pi^2 C_A}{2}+\frac{13 N_F}{18}\right)\right]
\ S_J^{(2)}(\vec q\,;x)\nonumber
\\
&+&
\left(\Lambda^{2}\right)^\varepsilon\int_0^1 d\beta
\left(P_{gg}(\beta)+2 N_F\frac{C_F}{C_A}P_{qg}(\beta)\right)
S_J^{(2)}(\vec q\,;x\beta) \nonumber
\\
&+&  2 \varepsilon  \int_0^1d\beta \left[ N_F \frac{C_F}{C_A} (1-\beta)\beta
\right.\nonumber \\
&+&\left.\left.
\, 2 \, C_A \left( \frac{\ln(1-\beta)}{1-\beta}
\right)_+(1-\beta)P(1-\beta)
\right]S_J^{(2)}(\vec q\,;x\beta) \right\}+\mathcal{O}(\varepsilon)\;.
\nonumber
\end{eqnarray}

Then, we collect the contributions given
in~(\ref{finitegC_Fa}),~(\ref{finitegC_Aa}),~(\ref{finiteA}),~(\ref{finiteB}),
and get (transforming to the form used in~\cite{bar2})
\begin{equation}
\label{vertexg}
V_g^{(2)}(\vec q\,)\equiv\left(V_g^{(R_{q\bar q})(C_F, {\rm finite})}
+V_g^{(R_{q\bar q})(C_A, {\rm finite})}+V_g^{(R_{gg})(A, {\rm finite})}
\right.
\end{equation}
\[
\left.
+V_g^{(R_{gg})(B, {\rm finite})}\right)(\vec q\,)
\]
\[
=
\int_0^1 dx f_g(x)\int_0^1 d\beta \left[
  2\, N_F \, P_{qg}(\beta)
\int\frac{d^{2}\vec k}{\pi \vec k^2}\left\lbrace
\frac{\vec q^{\: 2}}{\vec k^{2}+(\vec q-\vec k)^2}
S_J^{(3)}(\vec k,\vec q-\vec k,x\beta;x)\right.\right.
\]
\[
\left.
-\Theta(\Lambda^2-\vec k^2)
S_J^{(2)}(\vec q\,;x\beta)\right\}
\]
\[
+ \, N_F \frac{C_A}{C_F}\, P_{qg}(\beta) \int
\frac{d^{2}\vec k}{\pi (\vec k-(1-\beta)\vec q\,)^2}
\left\lbrace\frac{\vec q^{\: 2}\beta(1-\beta)\vec k\cdot
(\vec q\,-\vec k )}
{\vec k^{ 2}(\vec k-\vec q\,)^2} \right.
 S_J^{(3)}(\vec k,\vec q\, -\vec k,x\beta;x)
\]
\[
\biggl.
-\Theta(\Lambda^2-(\vec k-(1-\beta)\vec q\,)^2)S_J^{(2)}(\vec q\,;x)
\biggr\}\biggr]
\]
\[
+ \int_0^1 dx \, 2 \, C_A \frac{C_A}{C_F}\, f_g(x)\, \left[\int_0^1
\frac{d\beta}{(1-\beta)_+}[(1-\beta)P(1-\beta)]
\int \frac{d^{2} \vec l}{\pi \vec l^2}\right.
\]
\[
\times\left\lbrace
\frac{\vec q^{\: 2}}{\vec l^2+(\vec l-\vec q\,)^2}\left(\
S_J^{(3)}(\vec q-(1-\beta)\vec l,(1-\beta)\vec l,x(1-\beta);x)
\right.\right.
\]
\[
\left.
+ S_J^{(3)}(\beta \vec q+(1-\beta)\vec l,(1-\beta)(\vec q-\vec l\,),
x(1-\beta);x)\right)
-\biggl.\Theta\left(\Lambda^2-\vec l^{\:2}\,\right)
\left(S_J^{(2)}(\vec q\,;x\beta)
+S_J^{(2)}(\vec q\,;x)\!\right)\!\biggr\}
\]
\[
+ \int_0^1 d\beta
\int \frac{d^{2} \vec k}{\pi }\left\{P(\beta)\left(
\frac{\vec q^{\: 2}(1-\beta)\vec k \cdot(\vec k-\beta\vec q\,)}
{( \vec k-\beta \vec q\, )^2( \vec k-\vec q\, )^2 \vec k^{\, 2}}
\, S_J^{(3)}(\vec q-\vec k,\vec k,x\beta;x)\right.\right.
\]
\[
\left.\left. \left.
- \frac{1}{(\vec k-\vec q)^2}
\Theta \left(\Lambda^2-(\vec k-\vec q\,)^2 \right)
S_J^{(2)}(\vec q\,;x\beta)\right)-\frac{1}{\beta}\,\frac{\vec q^{\: 2}
\Theta[(1-\beta)|\vec q\, -\vec k|-\beta |\vec k|]}
{\vec k^2(\vec q-\vec k)^2}\, S_J^{(2)}(\vec k;x)\right\rbrace\right]\;.
\]

Besides, we define
\begin{eqnarray}
&&V_g^{(3)}(\vec q\,)\equiv\left(
V_g^{(R_{gg})(B, {\rm soft})}+V_g^{(C)}
\right)(\vec q\,)
\nonumber \\
&&
= \frac{C_A }
{(4\pi)^{\varepsilon}} \int_0^1 dx \, \frac{C_A}{C_F} \, f_g(x)
\int \frac{d^{D-2}\vec k}{\pi^{1+\varepsilon}}
\frac{\vec q^{\: 2}}{\vec k^{ 2}(\vec k-\vec q\,)^2}
\ln\frac{s_0}{(|\vec k|+|\vec q-\vec k|)^2} \ S^{(2)}_J(\vec q-\vec k;x)
\; ,
\end{eqnarray}
given in Eq.~(\ref{softCA+C_gluon}).

Another contribution originates from the collinear and charge renormalization
counterterms, see Eqs.~(\ref{charge.count.t}) and~(\ref{c.count.t.a}),
\[
V_g^{(4)}(\vec q\,)=\frac{\Gamma[1-\varepsilon]}
{\varepsilon \,(4\pi)^\varepsilon}
\int_0^1 dx\, f_g(x) \left[\left(\mu_R^2\right)^\varepsilon
\left(\frac{11C_A}{6}-\frac{N_F}{3} \right)\, \frac{C_A}{C_F}
\, S_J^{(2)}\left(\vec q\,;x\right)
\right.
\]
\begin{equation}
\left.
-
\left(\mu_F^{2}\right)^{\varepsilon} \int_0^1 d\beta\left[
2N_F P_{qg}(\beta)
+\frac{C_A}{C_F}P_{gg}(\beta)\right]
\ S^{(2)}_J(\vec q\,;x\beta)\right] \; .
\label{c.t.-g}
\end{equation}

Finally, the gluon part of the jet impact factor is given by the sum of the
above four contributions and can be presented as a sum of two terms:
\begin{equation}
\label{singular2g}
V_g^{(I)}(\vec q\,)=
\int_0^1 dx\, \frac{C_A}{C_F} \, f_g(x) \left[\frac{ C_A}
{(4\pi)^{\varepsilon}} \int \frac{ d^{D-2}\vec k}{\pi^{1+\varepsilon}}
\frac{\vec q^{\: 2}}{\vec k^{\, 2}(\vec k-\vec q\,)^2}
\ln\frac{s_0}{(|\vec k|+|\vec q-\vec k|)^2} \ S^{(2)}_J(\vec k;x)\right.
\end{equation}
\[
\left. -C_A\ln\left(\frac{s_0}
{\vec q^{\: 2}}\right) \left(\vec q^{\,\, 2}\right)^{\varepsilon}
\frac{\Gamma[1-\varepsilon]}{\varepsilon \,(4\pi)^\varepsilon}
\frac{\Gamma^2(1+\varepsilon)}{\Gamma(1+2\varepsilon)}
 S^{(2)}_J(\vec q\,;x) \right]
\]
and
\begin{equation}
V_g^{(II) }(\vec q\,)=V_g^{(2) }(\vec q\,)+
\int_0^1 dx\, \frac{C_A}{C_F} \, f_g(x)
\label{non-singular2g}
\end{equation}
\[
\times \left. \left\lbrace \left[  \left(
\frac{11 C_A}{6}-\frac{N_F}{3}\right)\,\ln\frac{\vec q^{\,\, 2}\mu_R^2}
{\Lambda^4} \, +\, C_A\frac{\pi^2}{2} +\frac{13}{18}N_F\right.\right.
\right]
\ S^{(2)}_J(\vec q\,;x)
\]
\[
+ \int_0^1 d\beta\left[P_{gg}(\beta)
+\,2\,N_F\,\frac{C_F}{C_A}P_{qg}(\beta)\right] \,\ln\frac{\Lambda^2}{\mu_F^2}
\, \ S^{(2)}_J(\vec q\,;x\beta)
\]
\[
+\left.
\int_0^1 d\beta\left[\,4\,\left(\frac{\ln(1-\beta)}{1-\beta}
\right)_+ [(1-\beta)P(1-\beta)] \,C_A + \,2\, N_F\,\frac{C_F}{C_A}
\beta(1-\beta)\right]\ S^{(2)}_J(\vec q\,;x\beta)\,
\right\rbrace\;.
\]

\section{Summary}

We have recalculated the jet vertices for the cases of quark and gluon in
the initial state, first found in the papers by Bartels 
{\it et al.}~\cite{bar1,bar2}. Our approach is different since the starting 
point of our calculation is the known general expression for NLO 
BFKL impact factors, given in Ref.~\cite{FF98}, applied to the special case of 
partons in the initial state. Nevertheless, in many technical steps we 
followed closely the derivation of Refs.~\cite{bar1,bar2}.

We checked our result by replacing everywhere the jet selection functions
$S_J^{(2)}$ and $S_J^{(3)}$ by 1 and performing all integrations: we got
back to known results for the NLO quark and gluon impact
factors~\cite{fadinq,fading}.

Let us discuss now the infrared finiteness of the obtained result for the jet
impact factor. The NLO correction to the jet vertex (impact factor) has the
form
\begin{equation}
\frac{d\Phi^{(1)}_J(\vec q\,)}
{dJ}\, = \, \frac{\alpha_s}{2\pi}\,  \Phi^{(0)}_q \, V (\vec q\,)\ ,\quad\quad
V(\vec q\,)=V^{(I)}(\vec q\,)+V^{(II)}(\vec q\,)\, ,
\end{equation}
where each part is the sum of the quark and gluon contributions,
$$ V^{(I)}(\vec q\,)=V^{(I)}_q(\vec q\,)+V^{(I)}_g(\vec q\,)\, , \quad
   V^{(II)}(\vec q\,)=V^{(II)}_q(\vec q\,)+V^{(II)}_g(\vec q\,)
$$
given in Eqs.~(\ref{singular2}),~(\ref{non-singular2}) and in
Eqs.~(\ref{singular2g}),~(\ref{non-singular2g}), respectively.
$V^{(II)}_q(\vec q\,)$ and $V^{(II)}_g(\vec q\,)$ are manifestly finite.
For $V^{(I)}(\vec q\,)$ we have
\begin{equation}
\label{singular2end}
V^{(I)}(\vec q\,)=
\int_0^1 dx\, \left( \sum_{a=q, \bar q} f_{a}(x)+ \frac{C_A}{C_F}
\, f_g(x) \right)
\end{equation}
\[
\times\left[\frac{ C_A}
{(4\pi)^{\varepsilon}} \int \frac{ d^{D-2}\vec k}{\pi^{1+\varepsilon}}
\frac{\vec q^{\: 2}}{\vec k^{\, 2}(\vec k-\vec q\,)^2}
\ln\frac{s_0}{(|\vec k|+|\vec q-\vec k|)^2} \ S^{(2)}_J(\vec k;x)\right.
\]
\[
\left. -C_A\ln\left(\frac{s_0}
{\vec q^{\: 2}}\right) \left(\vec q^{\,\, 2}\right)^{\varepsilon}
\frac{\Gamma[1-\varepsilon]}{\varepsilon \,(4\pi)^\varepsilon}
\frac{\Gamma^2(1+\varepsilon)}{\Gamma(1+2\varepsilon)}
 S^{(2)}_J(\vec q\,;x) \right] \, .
\]
Having the explicit form of the lowest order jet function~(\ref{jetF0}), it
is easy to see that the integration of $V^{(I)}(\vec q\,)$ over $\vec q$ with
any function, regular at $\vec q=\vec k_J$, will give some finite result.  In
particular, the finite result will be obtained after the convolution of
$V^{(I)}(\vec q\,)$ with BFKL Green's function, see Eq.~(\ref{crX}), which is
required for the calculation of the jet cross section.

This may look somewhat surprising, but, in fact, it is not so since the impact
factor is not, strictly speaking, a physical observable. The divergence
in~(\ref{singular2end}) arises from virtual corrections and, precisely, from
the factor $(s_0/\vec q^{\:2})^{\omega(\vec q^{\:2})}$ entering the
definition of the impact factor.  In the computation of physical impact
factors this divergence is cancelled by the one arising from the integration
in the first term of Eq.~(\ref{singular2end}), which is related with real
emission. In the calculation of the jet vertex the $\vec q$ integration is
``opened'' and, therefore, there is no way to get the divergence needed to
balance the one arising from virtual corrections. However, in the
construction of any physical cross section, the jet vertex is to be
convoluted with the BFKL Green's function, which implies the integration over
the Reggeon transverse momentum $\vec q$.

In our approach the energy scale $s_0$ remains untouched and need not
be fixed at any definite scale. The dependence on $s_0$ will disappear
in the next-to-leading logarithmic approximation in any physical cross
section in which jet vertices are used. However, the dependence on this
energy scale will survive in terms beyond this approximation and will
provide a parameter to be optimized with the method adopted in
Refs.~\cite{mesons,photons}.

In order to compare our results with the ones of~\cite{bar1,bar2}, we need to
perform the transition from the standard BFKL scheme with arbitrary energy
scale $s_0$ to the one used in~\cite{bar1,bar2}, where the scale of energy
depends on the Reggeon momentum. The change to the scheme where the energy
scale $s_0$ is replaced to any factorizable scale $\sqrt{f_1(\vec q_1^{\,\, 2})
f_2(\vec q_2^{\,\, 2})}$ leads to the following modification of each impact
factor ($i=1,2$), see~\cite{Fadin:1998sg},
\begin{equation}
\Phi_i(\vec q\,;f_i(\vec q^{\,\, 2}))=\Phi_i(\vec q\,;s_0)
+\frac{1}{2}\int d^{D-2}\vec k \ \Phi^{(0)}_i(\vec k)\,
\ln\left(\frac{f_i(\vec k^2)}{s_0}\right)K^{(0)}(\vec k,\vec q\, )
\ \frac{\vec q^{\: 2}}{\vec k^{\, 2}} \, ,
\label{change-scale}
\end{equation}
where $\Phi^{(0)}_i$ and $K^{(0)}$ are the lowest order impact factor and
BFKL kernel. Therefore changing from $s_0$ to $s_0=\sqrt{\vec q^{\, 2}_1
\vec q^{\, 2}_2}$ we obtain the following replacement in our result for the
jet impact factor:
\begin{equation}
\label{singular2F}
V^{(I)}(\vec q\,)\to \bar V^{(I)}(\vec q\,)=
\int_0^1 dx\, \left( \sum_{a=q, \bar q} f_{a}(x)+ \frac{C_A}{C_F}
\, f_g(x) \right)
\end{equation}
\[
\times\left[\frac{ C_A}
{(4\pi)^{\varepsilon}} \int \frac{ d^{D-2}\vec k}{\pi^{1+\varepsilon}}
\frac{\vec q^{\: 2}}{\vec k^{\, 2}(\vec k-\vec q\,)^2}
\ln\frac{\vec k^2}{(|\vec k|+|\vec q-\vec k|)^2} \ S^{(2)}_J(\vec k;x)\right]
\; .
\]
Note that $\bar V^{(I)}(\vec q\,)$ is not singular at $\vec q\to \vec k$ and,
therefore, it can be calculated at $D=4$. Such contribution to the jet
impact factors, $\bar V^{(I)}(\vec q\,)$, in the considered scheme with
$s_0=\sqrt{\vec q^{\, 2}_1\vec q^{\, 2}_2}$ produces a completely equivalent
effect on the physical jet cross section as the factors $H_L$ and $H_R$ which
enter Eq.~(76)
of~\cite{bar2} (see Eqs.~(101),~(102) in~\cite{bar1} for the definition of
$H_L$, $H_R$).

Therefore, for the final comparison one needs to consider our results for
$V^{(II)}_q(\vec q\,)$ and $V^{(II)}_g(\vec q\,)$ (modulo the appropriate
normalization factor) with the ones given in Eq.~(105) of~\cite{bar1} and
Eq.~(67)  of~\cite{bar2} for the quark and gluon contributions, respectively.
For this purpose we identify, following~\cite{bar1,bar2}, the $\Lambda$
parameter with the collinear factorization scale $\mu_F$. In the gluon
contribution we found a complete agreement, whereas in the quark case we point
out to the following misprints in the Eq.~(105) of~\cite{bar1}:
\begin{itemize}
\item  in the 'real' $C_A$ term, the expression {\bf $q$}$-${\bf $k$} must be
replaced by {\bf $q$}$- z${\bf $k$}, both in the numerator and the denominator;
\item $P_{qq}(z)$ in front of the same term has to be replaced by
${\cal P}_{gq}(0,z)$ or, similarly, by $P_{gq}(z)/C_F$;
\item just after it, $+-$ is to be interpreted as $-$.
\end{itemize}
Exactly these corrections were pointed out first in 
Ref.~\cite{Colferai:2010wu}. In that paper the results for the quark and 
gluon contributions to the jet vertices are presented in a form different 
from the one used in the original calculation of Refs.~\cite{bar1,bar2}.
However, one can see after some analysis, that these two forms turn to be 
completely equivalent, contrary to the impression that the text before 
Eq.~(3.4) in Ref.~\cite{Colferai:2010wu} may induce~\footnote{We are very 
thankful to the authors of Ref.~\cite{Colferai:2010wu} for the clarification 
of this point.}.

Recently another work devoted to the calculation of the jet impact factor
appeared~\cite{Hentschinski:2011tz}. It is an interesting application of the
Lipatov's effective action method~\cite{Lipatov:1995pn} to the problem in
question. Within this method, a particular regularization of the longitudinal
divergences has been proposed. In the traditional approach, these
divergences are regularized by the account of the BFKL counterterm, see
Eq.~(\ref{counterterm}). Unfortunately more work with the results
of~\cite{Hentschinski:2011tz}  seems  still to be done in order to get
an independent check of the final results for the jet impact factors.

\section*{Acknowledgements}

D.I. thanks the Dipartimento di Fisica dell'Universit\`a della Calabria
and the Istituto Nazio\-na\-le di Fisica Nucleare (INFN), Gruppo collegato di
Cosenza, for the warm hospitality and the financial support. This work was
also supported in part by the grants RFBR-09-02-00263 and RFBR-11-02-00242.
A.P. thanks A.~Sabio Vera for a very stimulating conversation.

\end{document}